\documentclass[twocolumn,pra,superscriptaddress,english,nofootinbib]{revtex4-2}
\usepackage{indentfirst}
\usepackage{ragged2e}
\usepackage{color}
\usepackage{amsmath}
\usepackage{amsfonts}
\usepackage{amssymb}
\usepackage{graphicx}
\usepackage[ruled, longend]{algorithm2e}

\usepackage[usenames,divipsnames,svgnames,table]{xcolor}
\usepackage[unicode=true,pdfusetitle,
bookmarks=true,bookmarksnumbered=false,bookmarksopen=false,
breaklinks=true,pdfborder={0 0 0},backref=false,colorlinks=true]{hyperref}
\hypersetup{linkcolor=NavyBlue,urlcolor=NavyBlue,citecolor=NavyBlue}
\usepackage{breakurl}

\begin{document}


\title{Optimal Scheme for Quantum Metrology}

\author{Jing Liu}
\email{liujingphys@hust.edu.cn}
\affiliation{MOE Key Laboratory of Fundamental Physical Quantities Measurement,
PGMF and School of Physics, Huazhong University of Science and Technology, Wuhan 430074, China}

\author{Mao Zhang}
\affiliation{MOE Key Laboratory of Fundamental Physical Quantities Measurement,
PGMF and School of Physics, Huazhong University of Science and Technology, Wuhan 430074, China}

\author{Hongzhen Chen}
\affiliation{Department of Mechanical and Automation Engineering, The Chinese University
of Hong Kong, Shatin, Hong Kong}

\author{Lingna Wang}
\affiliation{Department of Mechanical and Automation Engineering, The Chinese University
of Hong Kong, Shatin, Hong Kong}

\author{Haidong Yuan}
\email{hdyuan@mae.cuhk.edu.hk}
\affiliation{Department of Mechanical and Automation Engineering, The Chinese University
of Hong Kong, Shatin, Hong Kong}

\begin{abstract}
Quantum metrology can achieve far better precision than classical metrology, and is one of the most
important applications of quantum technologies in the real world. To attain the highest precision
promised by quantum metrology, all steps of the schemes need to be optimized, which include the
state preparation, parametrization and measurement. Here we review the recent progresses on the
optimization of these steps, which are essential for the identification and achievement of the
ultimate precision limit in quantum metrology. We hope this provides a useful reference for the
researchers in quantum metrology and related fields.
\end{abstract}

\maketitle

\section{Introduction}
\noindent

Metrology, which studies the precision limit of measurement and estimation, plays
a central role in science and technology. Recently quantum metrology, which
exploits quantum mechanical effects to achieve far better precision than classical
schemes~\cite{Helstrom1976,Holevo1982,Giovannetti2011,Giovannetti2006,anisimov2010quantum,
Braunstein1994,paris2009quantum,Fujiwara2008,Escher2011,Deng2021,demkowicz2014using,Rafal2012,
huelga1997improvement,chin2012quantum,HallPRX,Berry2015,Alipour2014,Beau2017,schnabel2010quantum,
Wang2017,Bai2019,Wu2021}, has gained increasing attention and has found wide applications
in various fields, such as gravitational wavedetection~\cite{schnabel2010quantum,Abadie2011,
Eberle2010,Steinlechner2018,Grote2013,Tse2019}, quantum phase estimation~\cite{anisimov2010quantum,
joo2011quantum,Higgins2007,Schafermeier2018,Oh2019}, quantum magnetometer~\cite{Wolfgramm2010,Li2018},
quantum ranging~\cite{Pooser2015,Huang2021,Zhuang2021}, quantum spectroscopy~\cite{Michael2019,Cai2021,
Liu2020,Li2020}, quantum imaging~\cite{kolobov1999spatial,lugiato2002quantum,Treps2003,morris2015imaging,
roga2016security,tsang2016quantum,Casacio2021}, quantum target-detection~\cite{shapiro2009quantum,
lopaeva2013experimental}, quantum gyroscope~\cite{dowling1998correlated,Che2019}, distributed quantum
sensing~\cite{Guo2020,Xia2020,Zhao2021}, atomic clocksynchronization~\cite{bollinger1996optimal,
Buzek1999,leibfried2004toward,roos2007designer,derevianko2011colloquium,ludlow2015optical,
borregaard2013near}, and even biological measurements~\cite{Taylor2013}.

A central task in quantum metrology is to identify the ultimate precision limit that can be
achieved with given resources and design schemes to attain it. In the finite regime where the number of
the measurement is limited, this is usually a hard task. When one does not have sufficient prior information
of the parameter, this is a `global' problem in the sense that the chosen cost function need to be minimized
over a certain region. This is often handled by minimizing the mean of the cost function or minimizing
the worst case over the region~\cite{Hayashi2006,Imai2009,Hayashi2011} for which the optimal
scheme is only known for very symmetric cases~\cite{Massar1995,Holevo1998,Chiribella2004,Chiribella2005,
Bisio2010}. The task is much simpler when the number of the measurement can be asymptotically large, in this
case the value of the parameter can be pinpointed asymptotically and the `global' problem becomes `local'.
In this regime, a well-behaviored cost function can be approximated by the variance, which corresponds to
the second order expansion of the cost function around the true value, and the local precision limit can be
quantified by the quantum Cram\'er-Rao bound~\cite{Helstrom1976, Holevo1982}. The studies on the optimal schemes
for the single-parameter estimation in the asymptotic limit have made much progress recently, which will
be the focus of this review. For the progresses on the multi-parameter estimation, we refer the
readers to the recent reviews~\cite{Liu2020,Rafal2020,Albarelli2020,Sidhu2020}.

\begin{figure}[tp]
\centering\includegraphics[width=9cm]{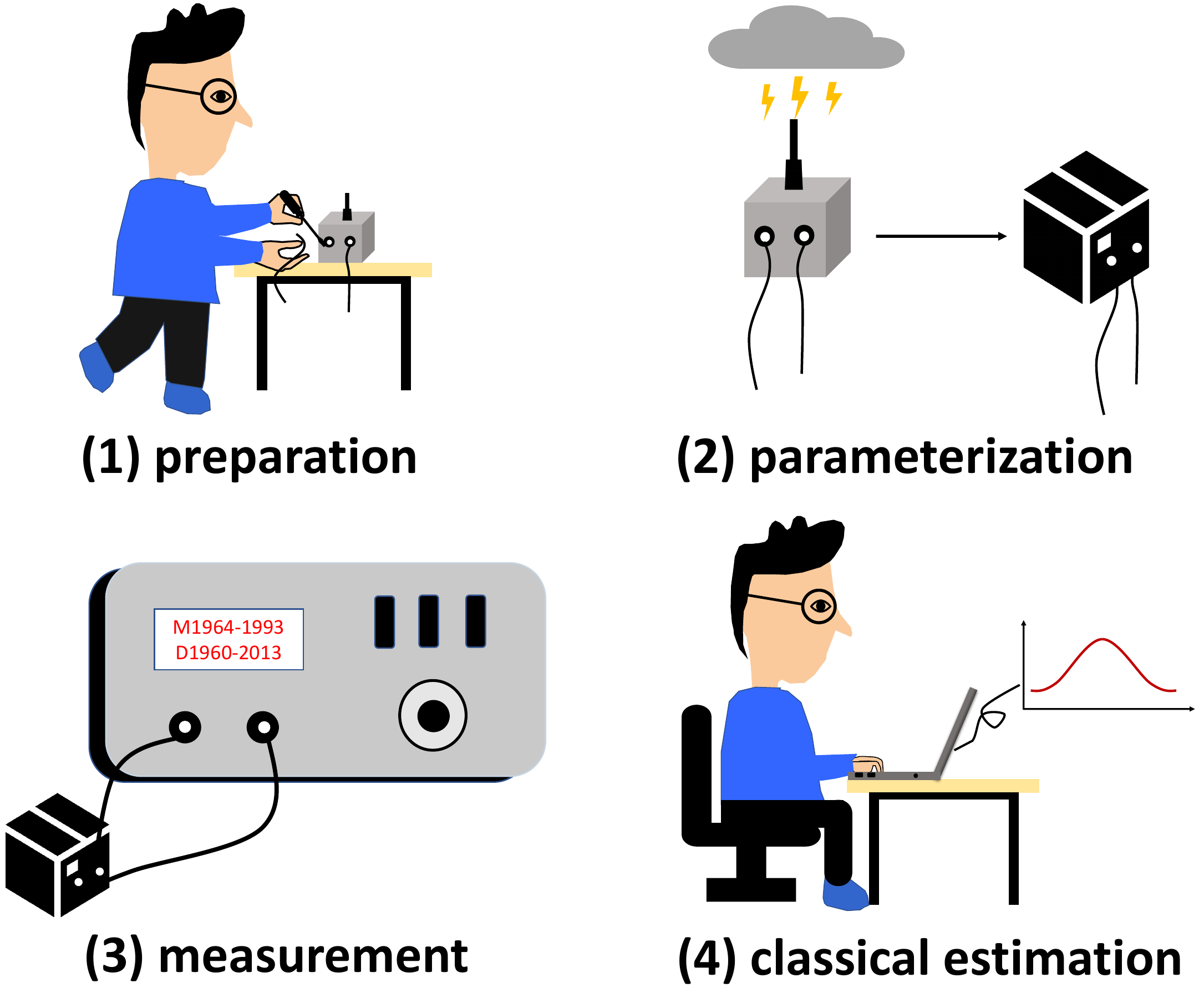}
\caption{The schematic for a general process of quantum parameter
estimation, which includes four steps: (1) preparation;
(2) parameterization; (3) measurement; and (4) classical estimation.
The first three steps (``ppm'' steps) are the major subjects in
quantum metrology. }
\label{fig:schematic_ppm}
\end{figure}

A general protocol of quantum metrology typically consists of four steps: (1) preparation;
(2) parameterization; (3) measurement and (4) estimation, as illustrated in Fig.~\ref{fig:schematic_ppm}.
To achieve the best precision, all steps need to be optimized. The optimization of the last step
is a well-studied subject in the classical statistics, in this review, we will focus on the optimization
of the first three steps (quantum part), denoted as ``ppm"\,\footnote{The initials of preparation,
parameterization and measurement.}. Interestingly, "ppm" is also one of the most well-used pseudo-units
in the field of precision measurements, standing for one part per million. In Sec.~\ref{sec:metrological_bounds}
we first review the quantum dynamics and the quantum Cram\'er-Rao bound, which are the main tools used
to quantify the precision limit of the single-parameter quantum estimation.
Section~\ref{sec:state_opt} to~\ref{sec:measurement_opt} focus on the current development on the
optimizations of the first three steps. Specifically, in Sec.~\ref{sec:state_opt}
we review the progresses on the optimization of the initial probe states. In Sec.~\ref{sec:parameterization_opt},
we review the progresses on the optimization of the parametrization, which we mainly focus on the quantum
control and quantum error correction. In Sec.~\ref{sec:measurement_opt} we review the optimization of the
measurement, mainly on the local adaptive measurements, which is practically easier to implement. We make
a summary in Sec.~\ref{sec:summary}.

\section{Quantum dynamics and metrological bounds}
\label{sec:metrological_bounds}

\subsection{Quantum dynamics}

Quantum dynamics is the foundation of quantum information, and even quantum
mechanics. For an isolated quantum system, the dynamics is depicted by the
Schr\"{o}dinger equation $\partial_t|\psi\rangle=-iH|\psi\rangle$ for quantum
state $|\psi\rangle$ or Liouville equation $\partial_t\rho=-i[H,\rho]$ for
density matrix $\rho$. Here $H$ is the Hamiltonian. The solution for this
equation is $\rho(t)=U(t)\rho_{\mathrm{in}}U^{\dagger}(t)$ with $U(t)=\mathcal{T}
\exp(-i\int^t_0H(\tau)\mathrm{d}\tau)$ a unitary operator. $\mathcal{T}$ is the
time-ordering operator. It reduces to $\exp(-iHt)$ if $H$ is time-independent.
Once the quantum noise is involved, the evolution is not unitary any more. Two
typical approaches to depict the noisy dynamics are the integral and differential methods.
The most widely used integral method is the Kraus representation, which can model general
completely positive and trace-preserving maps, and thereby a routine analytical representation
for the integrated quantum dynamics, which acts as quantum channels on the probe states. In
this representation, the output state of a quantum channel $\mathcal{E}(\cdot)$, with the initial
probe state as $\rho_{\mathrm{in}}$, can be described as
\begin{equation}
\mathcal{E}(\rho_{\mathrm{in}})=\sum^{m}_{j=1}K_j\rho_{\mathrm{in}}K^{\dagger}_j,
\label{eq:kraus_opt}
\end{equation}
where $K_j$ is a Kraus operator satisfying $\sum_j K^{\dagger}_j K_j=\openone$
with $\openone$ the identity operator. Note that given a quantum channel, the associated Kraus
operators are not unique. Any isometry, which satisfies $V^\dagger V=\openone_m$ with $V\in\mathbb{C}^{p\times m}$
($p\geq m$), can lead to an equivalent set of Kraus operators as $\tilde{K}_j=\sum_i v_{ji}K_i$
with $v_{ji}$ the $ji$th entry of $V$. Such equivalent Kraus operators can be obtained by first
appending $p-m$ zero operators as $\{K_1,\cdots, K_m, \bold{0},\cdots, \bold{0}\}$,
then making an equivalent unitary transformation on these expanded operators, $\tilde{K}_j=\sum_i v_{ji}K_i$,
here $v_{ji}$ are the entries of the first $m$ columns of a $p\times p$ unitary matrix
(the entries of the remaining $p-m$ columns do not appear as they act on the zero operators).

The dynamics can also be described with the differential equations. A widely used
form that describes the Markovian dynamics is the master equation
\begin{equation}
\partial_t \rho=-i[H,\rho]+\sum_{k}\left(\!\Gamma_k\rho\Gamma_k^\dagger
-\frac{1}{2}\{\Gamma_k^\dagger\Gamma_k,\rho\}\!\right)\!,
\label{eq:master_eq}
\end{equation}
where $\Gamma_k$ is the $k$th Lindblad operator. 

\subsection{Quantum Cram\'er-Rao bounds}

To estimate the value of a parameter, $x_{\mathrm{true}}$, from certain
amount of data that is sampled from a conditional probability distribution that depends on the
parameter---we use $p(y|x)$ to denote the conditional probability of event $y$ given $x$, the
Cram\'er-Rao bound provides an asymptotically achievable lower bound
on the variance of any locally unbiased estimators as~\cite{Cramer1946,Rao1945}
\begin{equation}
\mathrm{var}(\hat{x})\geq \frac{1}{nI_x},
\end{equation}
here $\hat{x}$ is a locally unbiased estimator which satisfies $E(\hat{x})=x_{\mathrm{true}}$,
$\frac{\mathrm{d}}{\mathrm{d}x}E(\hat{x})|_{x_{\mathrm{true}}}=1$, $\mathrm{var}(\hat{x})
=E[(\hat{x}-E(\hat{x}))^2]$ is the variance of the estimator, $n$ is the number of sampled
data and
\begin{equation}
I_x=\int_y \frac{[\partial_x p(y|x)]^2}{p(y|x)}\mathrm{d}y
\end{equation}
is the Fisher information~\cite{Fisher1925}. For multiple unknown parameters
\begin{equation}
\mathbf{x}=(x_0,x_1,\cdots,x_m,\cdots)^{\mathrm{T}},
\end{equation}
the counterpart of the variance is the covariance matrix $\mathrm{cov}(\hat{\mathbf{x}})$,
where the $jk$-th entry is given by
\begin{equation}
\left[\mathrm{cov}(\hat{\mathbf{x}})\right]_{jk}\!=\!\int_y p(y|x)
[\hat{x}_j\!-\!E(\hat{x}_j)][\hat{x}_k\!-\!E(\hat{x}_k)]\mathrm{d}y
\end{equation}
with $\hat{x}_{j(k)}$ as the locally unbiased estimator for $x_{j(k)}$. It is easy
to see that the diagonal entries of the covariance matrix are just the variances
for the corresponding parameters. The mutli-parameter Cram\'er-Rao bound is given by
\begin{equation}
\mathrm{cov}(\hat{\mathbf{x}})\geq \frac{1}{n}I_{\mathbf{x}}^{-1},
\end{equation}
where $I_{\mathbf{x}}$ is now the Fisher information matrix with the $jk$-th entry
given by
\begin{equation}
(I_{\mathbf{x}})_{jk}=\int_y\frac{\partial_{x_j}p(y|x)\partial_{x_k}p(y|x)}{p(y|x)}
\mathrm{d}y.
\end{equation}
The Cram\'er-Rao bound can be achieved asymptotically when $n\rightarrow\infty$.

In quantum parameter estimation, the data is obtained from the measurements on the
quantum state, $\rho(x)$, which depends on the unknown parameter. The most general
measurement in quantum mechanics is the positive operator-valued measure (POVM),
which can be described by a set of positive semidefinite operators, $\{\Pi_i\}$,
that satisfies $\sum_i\Pi_i=\bold{I}$. The probability of the measurement result, $p(i|x)$, which for simplicity we also denote as $p_i(x)$, can be obtained as
$p_i(x)=\mathrm{Tr}[\rho(x) \Pi_i]$. From which we can get the Fisher information
\begin{equation}
I_x=\sum_{i}\frac{[\partial_x p_i(x)]^2}{p_i(x)}.
\end{equation}
The quantum Fisher information (QFI)~\cite{Helstrom1976,Holevo1982}, denoted as $F_x$, corresponds to the maximal of the Fisher information over all POVMs,
\begin{equation}
F_x = \max_{\{\Pi_i\}}I_x(\{\Pi_i\}).
\end{equation}
This leads to the quantum Cram\'er-Rao bound as
\begin{equation}
\mathrm{var}\left(\hat{x}\right)\geq\frac{1}{nF_x},
\label{eq:QCRB}
\end{equation}
which is asymptotically achievable in the case of the single-parameter estimation.

For multiple parameters, the quantum Cram\'er-Rao bound is given by
\begin{equation}
\mathrm{cov}\left(\hat{\mathbf{x}}\right)\geq\frac{1}{n}\mathcal{F}^{-1},
\label{eq:QCRB_mul}
\end{equation}
where $\mathcal{F}$ is now the quantum Fisher information matrix (QFIM).  
The entries of the QFIM can be obtained as
\begin{equation}
\mathcal{F}_{jk}=\frac{1}{2}\mathrm{Tr}(\rho\left\{L_j, L_k\right\}),
\end{equation}
where $L_j$ is the symmetric logarithmic derivative (SLD) corresponding to the parameter $x_j$, which is the solution to the equation $\partial_{x_j}\rho=\frac{1}{2}\left(\rho L_j+L_j\rho\right)$, $\{L_j,L_k\}=L_jL_k+L_kL_j$ is the anti-commutator. In the case of
the single-parameter estimation, the QFIM reduces to the QFI, which is a scalar given by
\begin{equation}
F_x=\mathrm{Tr}\left(\rho L^2_x\right)
\end{equation}
where $L_x$ is the SLD for the single parameter $x$. For single-parameter estimation, the quantum Cram\'er-Rao bound can be saturated by the projective measurement on the eigenvectors of the SLD and the maximum likelihood estimations~\cite{Helstrom1976, Holevo1982,Giovannetti2011}, while the multi-parameter quantum Cram\'er-Rao bound is in general not attainable due to the incompatibility of the optimal measurements for different parameters.

The quantum Fisher information is closely related to the geometry of the quantum
states. In particular, it is closely related to the fidelity, an important concept
for the distinguishment of two quantum states. A widely used form of the fidelity
between two quantum states, $\rho_1$ and $\rho_2$, is $f(\rho_1,\rho_2):=\mathrm{Tr}
\sqrt{\sqrt{\rho_1}\rho_2\sqrt{\rho_1}}$~\cite{Wootters1981,Nielsen2002}. This
leads to a distance measure on quantum states as
\begin{equation}
D(\rho_1,\rho_2):=\sqrt{2-2f(\rho_1,\rho_2)},
\end{equation}
which is referred to as the Bures distance. The fidelity can also be used to define
the Bures angle as $\Theta(\rho_1,\rho_2)=\arccos f(\rho_1,\rho_2)$, which is also a valid metric on the state space.
The QFI is proportional to the second order expansion of the Bures
distance~\cite{Braunstein1994}, i.e., up to the second order of $dx$
\begin{equation}\label{eq:Bures}
D(\rho_x,\rho_{x+\mathrm{d}x})=\frac{1}{4}F_x(\rho_x)
\mathrm{d}^2x,
\end{equation}
which shows that the QFI is nothing but the
fidelity susceptibility (ignoring the constant). A rigorous proof of this relation
for arbitrary-rank density matrices can be found in Refs.~\cite{Liu2020,Liu2014}
when the support is fixed.

The connection between the fidelity and the Fisher information can be generalized to quantum channels~\cite{Yuan2017c}.
Specifically given two quantum channels(the number of the Kraus operators for both channels, denoted as $m$, are assumed to be the same as we can always append zero operators if they are not the same initially),
\begin{eqnarray}
\aligned
\mathcal{E}_{1}(\rho)&=\sum_{i=1}^m K_{1,i}\rho K_{1,i}^\dagger, \\
\mathcal{E}_2(\rho)&=\sum_{i=1}^m K_{2,i}\rho K_{2,i}^\dagger,
\endaligned
\end{eqnarray}
the
fidelity between two channels is given by~\cite{Yuan2017c}
\begin{equation} \label{eq:f_qc}
f_{\mathrm{qc}}(\mathcal{E}_{1},\mathcal{E}_{2})=\max_{\Vert W\Vert_{\mathrm{op}}
\leq 1}\frac{1}{2}\lambda_{\min}(\mathcal{K}+\mathcal{K}^{\dagger}),
\end{equation}
where $\lambda_{\min}(\mathcal{K}+\mathcal{K}^{\dagger})$ denotes the minimum eigenvalue
of $\mathcal{K}+\mathcal{K}^{\dagger}$ with $\mathcal{K}=\sum_{ij}w_{ij}K^\dagger_{1,i}
K_{2,j}$, $w_{ij}$ is the $ij$th entry of any matrix, $W$, that satisfies
$\|W\|_{\mathrm{op}}\leq 1$ with $\|\cdot\|_{\mathrm{op}}$ denoting the operator
norm which equals to the largest singular value. Here $W$ comes from the non-uniqueness
of the Kraus representation for a quantum channel. Basically, given the equivalent
Kraus operators, $\tilde{K}_{1,q}=\sum_i [V_1]_{qi}K_{1,i}$ and $\tilde{K}_{2,q}=\sum_i [V_2]_{qi}K_{2,i}$,
with $V_1$, $V_2\in \mathbb{C}^{p\times m}$ as isometries, we have $\sum_q\tilde{K}^\dagger_{1,q}
\tilde{K}_{2,q}=\sum_{ij}w_{ij}K^\dagger_{1,i}K_{2,j}$ where $w_{ij}$ are entries of $W=V_1^\dagger V_2$ with
$\|W\|_{\mathrm{op}}\leq\|V_1^\dagger\|_{\mathrm{op}}\|V_2\|_{\mathrm{op}}=1$, i.e.,
\begin{eqnarray*}
& & \max_{\Vert W\Vert_{\mathrm{op}}
\leq 1}\frac{1}{2}\lambda_{\min}(\mathcal{K}+\mathcal{K}^{\dagger}) \\
&=& \max_{\{\tilde{K}_{1,q}\},\{\tilde{K}_{2,q}\}}\frac{1}{2}\lambda_{\min}
\left(\sum_q\tilde{K}^\dagger_{1,q}\tilde{K}_{2,q}+\tilde{K}^\dagger_{2,q}
\tilde{K}_{1,q}\right)\!\!.
\end{eqnarray*}
The fidelity on quantum channels can be efficiently computed through the semidefinite
programming as
\begin{eqnarray} \label{eq:fidelitySDP}
f_{\mathrm{qc}}  =& \max~\frac{1}{2}y, \nonumber \\
& \text{s.t.}~\begin{cases}
\left(\begin{array}{cc}
\openone & W^{\dagger}\\
W & \bold{I}
\end{array}\right) \geq 0, \\
\mathcal{K}+\mathcal{K}^{\dagger}-y\bold{I}\geq 0.
\end{cases}
\end{eqnarray}

Similar as the Bures distance on quantum states, we can define the Bures distance
on quantum channels as
\begin{equation}
D_{\mathrm{qc}}(\mathcal{E}_{1}, \mathcal{E}_{2})=\sqrt{2-2f_{\mathrm{qc}}
(\mathcal{E}_{1},\mathcal{E}_{2})}.
\end{equation}
The Bures distance on quantum channels corresponds to the minimal operator distance
among the equivalent Kraus
operators of the quantum channels as
\begin{eqnarray}
& & \min_{\{\tilde{K}_{1,q}\},\{\tilde{K}_{2,q}\}}\!\left\|\sum_q(\tilde{K}_{1,q}
-\tilde{K}_{2,q})^{\dagger}(\tilde{K}_{1,q}-\tilde{K}_{2,q})\right\|_{\mathrm{op}} \nonumber \\
&=& 2-\!\max_{\{\tilde{K}_{1,q}\},\{\tilde{K}_{2,q}\}}\!\lambda_{\min}\!
\left(\!\sum_q\tilde{K}^\dagger_{1,q}\tilde{K}_{2,q}+\tilde{K}^\dagger_{2,q}
\tilde{K}_{1,q}\!\right) \nonumber \\
&=& 2-2f_{\mathrm{qc}}(\mathcal{E}_{1},\mathcal{E}_{2}).
\end{eqnarray}

The quantum channels can be equivalently represented as
$\mathcal{E}_i(\rho_{\mathrm{S}})=\mathrm{Tr}_{\mathrm{E}}(U_{\mathrm{ES},i}(|0_{\mathrm{E}}\rangle
\langle 0_{\mathrm{E}}|\otimes\rho_{\mathrm{S}}) U^{\dagger}_{\mathrm{ES},i})$,
where $|0_{\mathrm{E}}\rangle$ denotes some standard state of the environment,
and $U_{\mathrm{ES},i}$ is a unitary operator acting on both the system and
environment, which is called the unitary extension of $\mathcal{E}_i$.
Similar to the Uhlmann's purification on quantum
states~\cite{Uhlmann1976}, the fidelity function on quantum channels also satisfies
\begin{eqnarray}
f_{\mathrm{qc}}(\mathcal{E}_1, \mathcal{E}_2)&=&\max_{U_{\mathrm{ES},1}}f_{\mathrm{qc}}
(U_{\mathrm{ES},1}, U_{\mathrm{ES,}2}) \\
&=&\max_{U_{\mathrm{ES},2}}f_{\mathrm{qc}}(U_{\mathrm{ES},1}, U_{\mathrm{ES},2}).
\end{eqnarray}

Operationally the fidelity between quantum channels equals to the minimal
fidelity between the output states of the extended channels as~\cite{Yuan2017c,Belavkina2005,Yuan2017b}
\begin{equation}
f_{\mathrm{qc}}(\mathcal{E}_1, \mathcal{E}_2)=\min_{\rho_{\mathrm{SA}}}
f[\mathcal{E}_1\otimes \bold{I} (\rho_{\mathrm{SA}}), \mathcal{E}_2\otimes\bold{I}
(\rho_{\mathrm{SA}})],
\label{eq:fidelitystate}
\end{equation}
where $\rho_{\mathrm{SA}}$ denotes a state on the system+ancilla,
and $\bold{I}$ is the identity operator on the ancillary system. This can be seen as an extension of the operational meaning of the fidelity between quantum states, which equals to the minimal classical fidelity between the probability distributions of the measurement result as
\begin{equation}
f(\rho_1,\rho_2)=\min_{\{E_i\}}f_{\mathrm{cl}}(p_1,p_2),
\end{equation}
where $f_{\mathrm{cl}}(p_1,p_2)=\sum_i\sqrt{p_{1,i}p_{2,i}}$ is the classical
fidelity with $p_{1,i}=\mathrm{Tr}(\rho_1\Pi_i)$ and $p_{2,i}=\mathrm{Tr}(\rho_2\Pi_i)$.
It is also known as Bhattacharyya coefficient~\cite{Bhattacharyya1943}.

\begin{figure}[tp]
\centering\includegraphics[width=9cm]{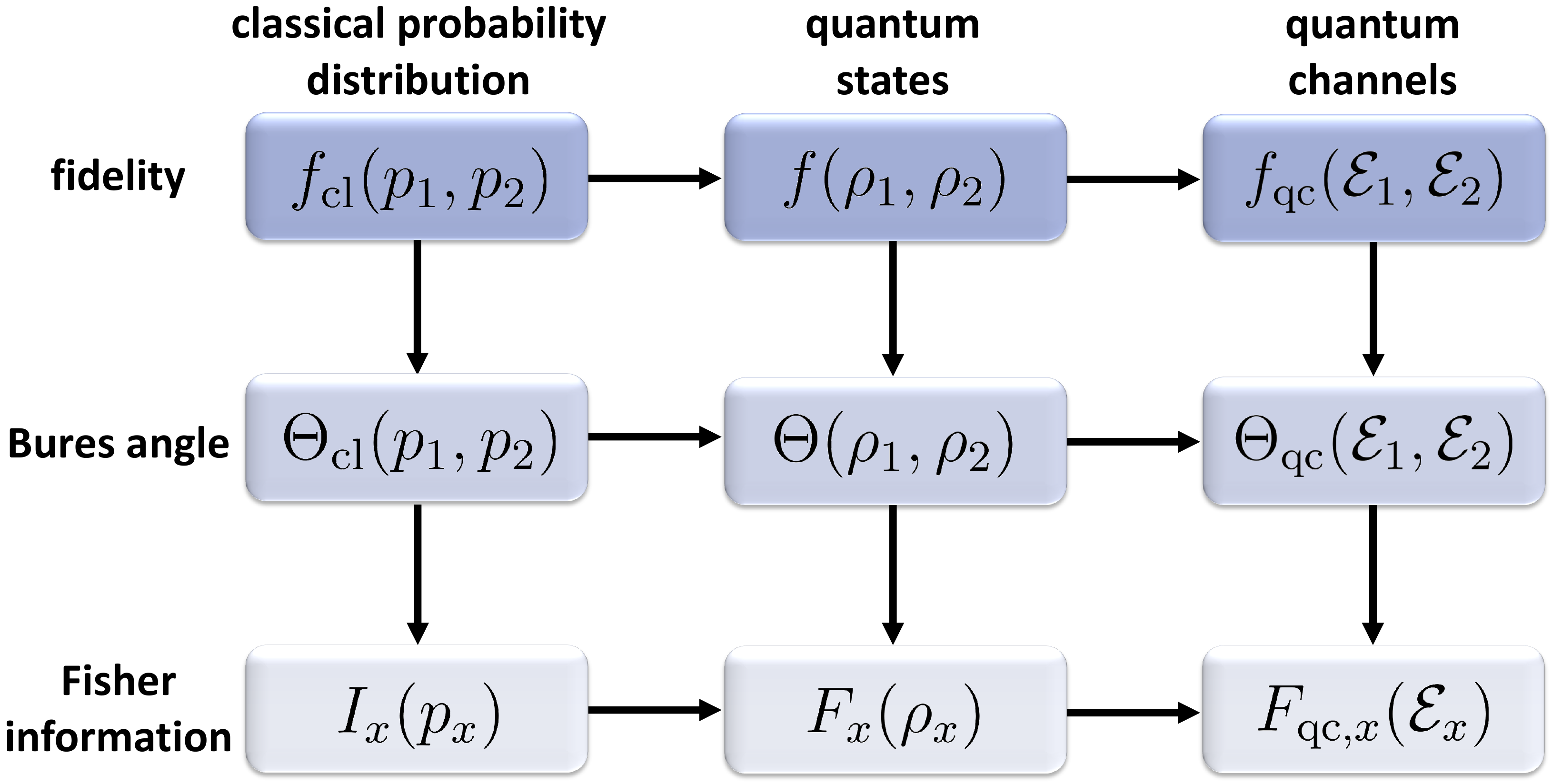}
\caption{A unified framework for the discriminations of
classical probability distributions, quantum states and
quantum channels, here $f_{\mathrm{cl}}(p_1,p_2)=\sum_i \sqrt{p_{1i}p_{2i}}$ with
$p_{1i}=\mathrm{Tr}(\rho\Pi_i)$ and $p_{2i}=\mathrm{Tr}(\rho_2\Pi_i)$,
$f(\rho_1,\rho_2)=\mathrm{Tr}(\sqrt{\rho_1^{\frac{1}{2}}\rho_2\rho_1^{\frac{1}{2}}})$,
and $f_{\mathrm{qc}}=\max_{\Vert W\Vert_{\mathrm{op}}\leq 1}\frac{1}{2}
\lambda_{\min}(\mathcal{K}+\mathcal{K}^{\dagger})$ as given in Eq.~(\ref{eq:f_qc}),
$\Theta_{\mathrm{cl}}=\arccos f_{\mathrm{cl}}$, $\Theta=\arccos f$,
$\Theta_{\mathrm{qc}}=\arccos f_{\mathrm{qc}}$, and the corresponding Fisher
information $I_x$, $F_x$ and $F_{\mathrm{qc},x}$ are given in Eqs.~(\ref{eq:Ix}),
(\ref{eq:Fx}) and (\ref{eq:QFI_qc}), respectively. The quantities associated with
the quantum states equal to the optimal value of the corresponding quantities on
the classical probability distribution after the optimization of the measurement,
and the quantities on quantum channels equal to the optimal value of the
corresponding quantities on quantum states after the optimization of the initial
probe states.}
\label{fig:QFI_relation}
\end{figure}

The Bures angle can also be extended to the quantum channels as
\begin{eqnarray}
\Theta_{\mathrm{qc}}(\mathcal{E}_1, \mathcal{E}_2)&=&\arccos f_{\mathrm{qc}}
(\mathcal{E}_1, \mathcal{E}_2).
\end{eqnarray}
The corresponding quantum channel Fisher information, which is the maximal
quantum Fisher information over all input states of the extended channel $\mathcal{E}_{x}\otimes  \bold{I}$, can
be related to the Bures angle as
\begin{equation}
F_{\mathrm{qc},x}=\lim_{\delta x\rightarrow 0}\frac{4\Theta^2_{\mathrm{qc}}
(\mathcal{E}_{x},\mathcal{E}_{x+\delta x})}{\delta x^2}.
\label{eq:QFI_qc}
\end{equation}
We then have a hierarchy of the Fisher information as
\begin{eqnarray}
I_x(p_x) &=& \lim_{\delta x\rightarrow 0}\frac{4\Theta^2_{\mathrm{cl}}(p_{x},
p_{x+\delta x})}{\delta x^2},
\label{eq:Ix}\\
F_x(\rho_x) &=& \lim_{\delta x\rightarrow 0}\frac{4\Theta^2(\rho_{x},
\rho_{x+\delta x})}{\delta x^2},
\label{eq:Fx}\\
F_{\mathrm{qc},x}(\mathcal{E}_{x})&=&\lim_{\delta x\rightarrow 0}\frac{4\Theta^2_{\mathrm{qc}}
(\mathcal{E}_{x},\mathcal{E}_{x+\delta x})}{\delta x^2}.
\end{eqnarray}
where $\Theta_{\mathrm{cl}}(p_x,p_{x+\delta x})=\arccos f_{\mathrm{cl}}(p_x,p_{x+\delta x})$
can be interpreted as the Bures angle between classical probabilities and
$\Theta(\rho_x,\rho_{x+\delta x})=\arccos f(\rho_x,\rho_{x+\delta x})$ is the
Bures angle between quantum states. This is summarized in Fig.~\ref{fig:QFI_relation},
in which the quantities on quantum states equal to the corresponding quantities on
the probability distribution over the optimization of the measurement and the
quantities on the quantum channels equal to the corresponding quantities on
quantum states over the optimization of the input states.

This is based on the purification approach developed in~\cite{Fujiwara2008,
 Escher2011,demkowicz2014using,Rafal2012,Sarovar2006,
Kolodynski2013}, which focus on the neighboring channels that are relevant
to the local precision limit. Given a quantum channel $\mathcal{E}_x(\rho)=\sum_i K_i(x)
\rho K_i^\dagger(x)$, the QFI, in terms of the initial
probe state and the Kraus operators, can be written as
\begin{equation}
F_x=\min_{\{\tilde{K}_j\}} C_{\mathrm{F}}(\rho_{\mathrm{in}},\{\tilde{K}_j\}),
\end{equation}
where $C_{\mathrm{F}}=4(\langle G_1\rangle_{\mathrm{in}}-\langle G_2\rangle^2_{\mathrm{in}})$,
$\rho_{\mathrm{in}}$ is the input probe state of the system (or the system+ancilla)
and $\langle G_{1(2)}\rangle_{\mathrm{in}}:=\mathrm{Tr}(G_{1(2)}\rho_{\mathrm{in}})$
with
\begin{equation}
G_1\!=\!\sum_j (\partial_x \tilde{K}^{\dagger}_j)(\partial_x \tilde{K}_j), \,
G_2=i\sum_j (\partial_x \tilde{K}^{\dagger}_j)\tilde{K}_j.
\label{eq:G_12}
\end{equation}
Here $\{\tilde{K}_j\}$ is a set of equivalent Kraus operators of the channel.
The channel Fisher information, after optimizing $\rho_{\mathrm{in}}$ on the
system+ancilla, is given by
\begin{equation} \label{eq:purificationQFI}
F_{\mathrm{qc},x}=\max_{\{\tilde{K}_j\}}\|G_1\|_{\mathrm{op}}.
\end{equation}
This expression of the channel QFI is equivalent to the channel QFI in
Eq.~(\ref{eq:QFI_qc})~\cite{Yuan2017b}. By restricting the equivalent transformation
of the Kraus operators to unitary transformation, i.e., $\tilde{K}_j=\sum_{p}v_{jp}K_p$
with $V$ (with the $jp$-th entry as $v_{jp}$)
as unitary instead of isometry, alternative semi-definite programming  that computes the channel QFI in
Eq.~(\ref{eq:purificationQFI}) has also been developed~\cite{Rafal2012,Kolodynski2013}.
If the channel is used $N$ times, it can be shown that for any input state, the QFI is
upper bounded by~\cite{Fujiwara2008,Escher2011,demkowicz2014using,Rafal2012,Kolodynski2013}
\begin{eqnarray} \label{eq:Nupperbound}
\aligned
4\min_{\{\tilde{K}_j\}}\,&[N\Vert G_1\Vert_{\mathrm{op}}
\\&+N(N-1)\Vert G_2\Vert_{\mathrm{op}}(\Vert G_1\Vert_{\mathrm{op}}+\Vert G_2\Vert_{\mathrm{op}}+1)].
\endaligned
\end{eqnarray}
The precision then cannot achieve the Heisenberg limit if there exists $\{\tilde{K}_j\}$
such that $G_2=0$.

\section{State optimization}
\label{sec:state_opt}

The state optimization is the first step in the design of optimal schemes for
quantum metrology. In the asymptotic regime, a natural target function
for the optimization is the QFI, or the CFI when the measurement is fixed.
Both analytical and numerical methods have been used in the state optimization. If the system is
simple or possesses some special properties that can facilitate the optimization,
the analytical approaches can be applied. For general complex systems numerical methods typically need to be employed.

Given a quantum channel,
$\mathcal{E}_{x}$, the optimal state that achieves the channel QFI, or equivalently
the optimal state that saturates the channel fidelity, $f_{\mathrm{qc}}(\mathcal{E}_{x},
\mathcal{E}_{x+\delta x})$, can be obtained through the dual optimization of the
problem in Eq.~(\ref{eq:fidelitySDP})~\cite{Yuan2017a}. Specifically, given  $\mathcal{E}_{x}(\rho)=\sum_{i=1}^mK_i(x)
\rho K_i^\dagger(x)$, we can define a quantum metrology matrix, $M$, which is a
$m\times m$ matrix with the $ij$th entry given by
\begin{equation}
M_{ij}(x_1,x_2):=\langle \psi_{\mathrm{SA}}| K_i(x_1)K_j(x_2)|\psi_{\mathrm{SA}}\rangle,
\end{equation}
where $|\psi_{\mathrm{SA}}\rangle$ is
the initial probe state of the system+ancilla and
\begin{equation*}
\langle\psi_{\mathrm{SA}}| K_i(x_1)K_j(x_2)|\psi_{\mathrm{SA}}\rangle
=\mathrm{Tr}[\rho_{\mathrm{s}}K_i(x_1)K_j(x_2)]
\end{equation*}
with $\rho_{\mathrm{s}}=\mathrm{Tr}_{\mathrm{A}}(|\psi_{\mathrm{SA}}\rangle\langle\psi_{\mathrm{SA}}|)$.
Given any pure input state the fidelity between the output states of two extended channel,
$\rho_{x_1}=\mathcal{E}_{x_1}\otimes \bold{I}(|\psi_{\mathrm{SA}}\rangle\langle\psi_{\mathrm{SA}}|)$ and
$\rho_{x_2}=\mathcal{E}_{x_2}\otimes \bold{I}(|\psi_{\mathrm{SA}}\rangle\langle\psi_{\mathrm{SA}}|)$,
equals to the trace norm of the quantum metrology matrix, i.e.,
\begin{equation}
f(\rho_{x_1},\rho_{x_1})=\mathrm{Tr}\sqrt{\sqrt{\rho_{x_1}}\rho_{x_2}\sqrt{\rho_{x_1}}}
=\left\Vert M(x_1,x_2)\right\Vert_{\mathrm{tr}},
\end{equation}
here $\left\Vert M(x_1,x_2)\right\Vert_{\mathrm{tr}}:=\mathrm{Tr}\sqrt{M M^{\dagger}}$,
which equals to the sum of singular values of $M$. The QFI of the output state
can then be obtained as
\begin{equation}
F_x = \lim_{\delta x\rightarrow 0}\frac{8\left(1-\left\Vert M(x,x+\delta x)
\right\Vert_{\mathrm{tr}}\right)}{\delta x^2}.
\label{eq:MF}
\end{equation}
The trace norm of a matrix can be efficiently calculated via the semidefinite
programming as~\cite{Yuan2017a}
\begin{eqnarray}
\left\Vert M\right\Vert_{\mathrm{tr}}&=&\min~\frac{1}{2}\mathrm{Tr}(P+Q), \nonumber \\
& & \mathrm{s. t.}\,\left(\begin{array}{cc}
P & M^{\dagger}\\
M & Q
\end{array}\right)\geq0,
\end{eqnarray}
where $P$, $Q$ are two Hermitian matrices. With the above formulation, the optimal state
can then be obtained by minimizing $\left\Vert M\right\Vert_{\mathrm{tr}}$ through the
following semi-definite programming
\begin{eqnarray}\label{eq:M}
\min_{\rho_{\mathrm{s}}}\left\Vert M\right\Vert_{\mathrm{tr}}
&=&\min~\frac{1}{2}\mathrm{Tr}(P+Q), \nonumber \\
& &\mathrm{s.t.}~\begin{cases}
\left(\begin{array}{cc}
P & M^{\dagger}\\
M & Q
\end{array}\right) \geq 0, \\
\rho_{\mathrm{s}} \geq 0, \\
\mathrm{Tr}(\rho_{\mathrm{s}}) = 1.
\end{cases}
\end{eqnarray}
Any $|\psi_{\mathrm{SA}}\rangle$
that has the reduced state equals to $\rho_{\mathrm{s}}$ outputed from the semi-definite
programming is an optimal probe state. This semi-definite programming is exactly the dual of the
semi-definite programming in Eq.~(\ref{eq:fidelitySDP}) and strong duality holds, the optimal values of both semi-definite programming give the channel fidelity.

The fact that the initial state enters $M$
linearly is essential for the formulation of the semi-definite programming in Eq.~(\ref{eq:M}). 
As a contrast, if the fidelity is calculated directly as
\begin{equation}
f=\mathrm{Tr}\sqrt{\sqrt{\rho_{x_1}}\rho_{x_2}
\sqrt{\rho_{x_1}}}
\end{equation}
with $\rho_x=\mathcal{E}_{x}\otimes \bold{I}(|\psi_{\mathrm{SA}}\rangle\langle\psi_{\mathrm{SA}}|)$,
the initial state then enters in quadratically,  which cannot be computed directly with the semi-definite
programming.

The formula in Eq.~(\ref{eq:MF}) also holds without the ancillary system when the
input state is a pure state, $|\psi_{\mathrm{s}}\rangle$. If the optimal value in Eq.~(\ref{eq:M}) can be attained with a pure $\rho_{\mathrm{s}}$, it then indicates that the channel QFI can be achieved
without the ancillary system, i.e., the ancillary system does not help improve the
precision in this case. This happens if all the operators $K_i^{\dagger}(x_1)K_j(x_2)$
commute with each other~\cite{Yuan2017a}, i.e.,
\begin{equation}
\left[K_i^{\dagger}(x_1)K_j(x_2), K_p^{\dagger}(x_1)K_l(x_2)\right]=0
\end{equation}
for any subscript $i,j,p,l$, for such channels the optimal value in Eq.~(\ref{eq:M}) can always be attained with a pure $\rho_{\mathrm{s}}$~\cite{Yuan2017a}, the ancillary system thus does not help
improve the precision limit. In particular, the unitary channel, the phase estimation with the
dephasing noise along the same direction or the phase estimation along the Z
direction with the noises along the X and Y directions, all satisfy this condition,
hence the ancillary system does not help improve the precision limit for these
channels in the case of single-parameter estimation~\cite{Yuan2017a}. For example, for the
dephasing channel
\begin{equation}
\mathcal{E}_x(\cdot)=K_1(\cdot)K_1+K_2(\cdot)K_2
\end{equation}
with $K_1=\sqrt{p}e^{-ix\sigma_z}$ and $K_2=\sqrt{1-p}\sigma_ze^{-ix\sigma_z}$, it is easy to check that
\begin{equation}
\left[K_i^{\dagger}(x_1)K_j(x_2), K_p^{\dagger}(x_1)K_l(x_2)\right]=0
\end{equation}
for any subscript $i,j,p,l\in\{1,2\}$, the ancillary system thus can not help improving the precision
limit in this case. This example is first recognized in~\cite{demkowicz2014using} through direct comparison.

When the dimension of the system gets large, the semi-definite programming becomes
computationally hard. It is then difficult to obtain the optimal state in the
general case. However, for some special case, it is still possible to obtain the
optimal state analytically.

\subsection{Analytical optimization}

Analytical optimizations are difficult to perform in most cases. In general the
complexity for the calculation of the QFI is equivalent to the diagonalization
of the density matrix, which grows exponentially with the number of particles.
However, there still exist some cases that the analytical optimization is possible.
The simplest case is the unitary parameterization with $\exp(-ixH)$, where the
optimal state can be analytically obtained~\cite{Giovannetti2006} as the
equal superposition of two eigenstates that corresponds to the maximal and minimal
eigenvalues of $H$ respectively.

\begin{figure}[tp]
\centering\includegraphics[width=9cm]{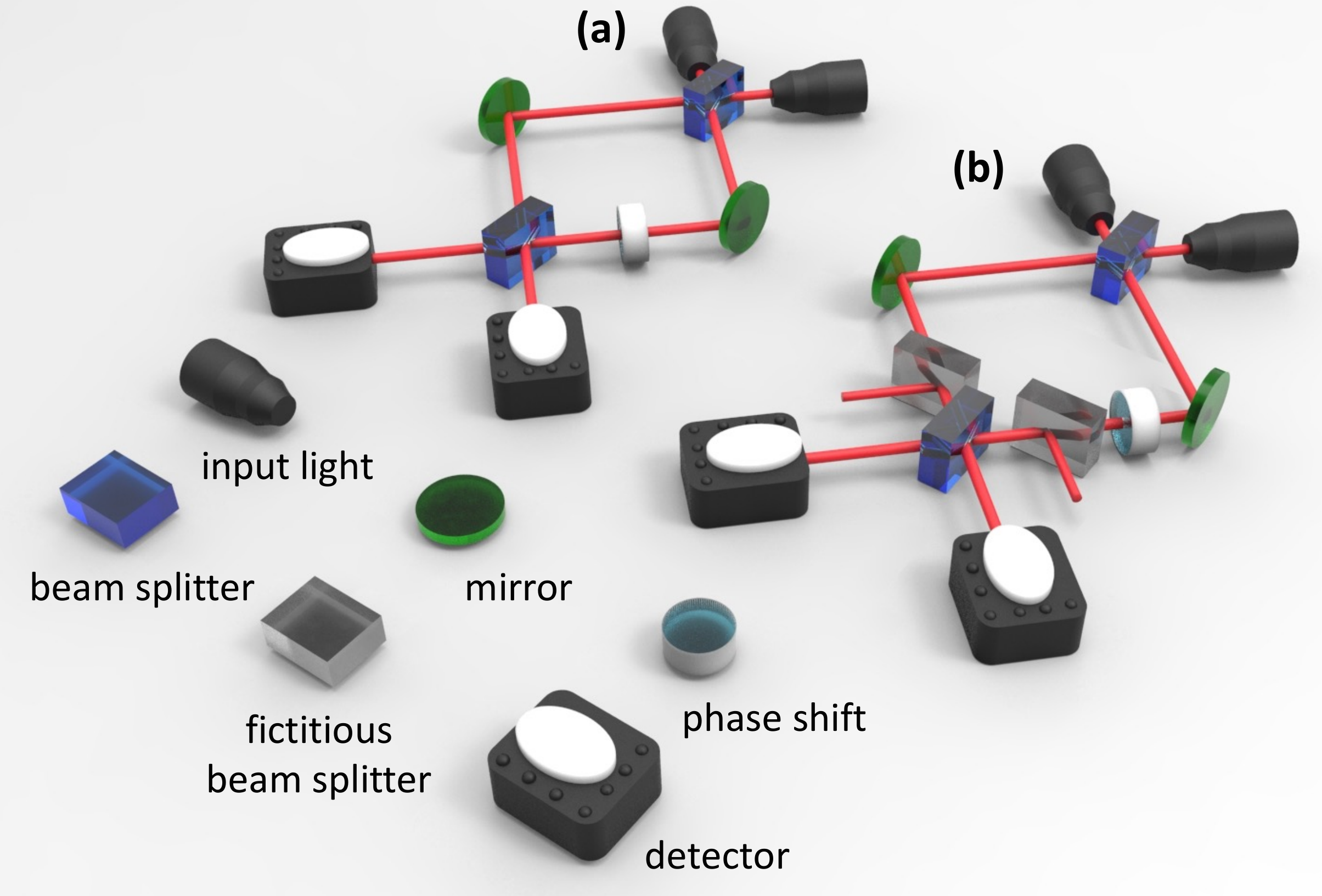}
\caption{(a) Scheme for a general optical Mach-Zehnder
interferometer for quantum phase estimation, which consists of
two beam splitters and a phase shift. (b) Scheme
for the lossy Mach-Zehnder interferometer in which the photon loss
in each arm is modeled by a fictitious beam splitter.}
\label{fig:MZI}
\end{figure}

The linear Mach-Zehnder interferometer (MZI) is another important scenario that
the analytical state optimization has been studied. An optical MZI consists of
two beam splitters and a phase shift as illustrated in Fig.~\ref{fig:MZI}(a).
This can be modeled as the SU(2) interferometer, of which the parameterization
process is $U_{\mathrm{mz}}=\exp(-i\phi J_y)$, where $J_y=\frac{1}{2i}(a^{\dagger}b
-ab^{\dagger})$ is a Schwinger operator with $a$ ($a^{\dagger}$) and $b$ ($b^{\dagger}$)
as the annihilation (creation) operators for two boson modes respectively
(the other two Schwinger operators are $J_x=\frac{1}{2}(a^{\dagger}b+ab^{\dagger})$
and $J_z=\frac{1}{2}(a^{\dagger}a-b^{\dagger}b)$~\cite{Caves1981}. In practice,
the input states in the two modes are typically separable, which significantly
reduce the state space for the optimization. This space can be further reduced
when additional restrictions are invoked. For example, if the probe state in one
import is restricted to be an odd or even state, then the maximal QFI can be
achieved when the states of the two modes satisfy the phase-matching condition~\cite{Liu2013},
which forms a basis for the state optimization in the MZI~\cite{Liu2013,Yu2018,Liang2020}.
In practice, the phase of the input states require extra resources to identify,
such as an external phase reference. Without these extra resources, the phase
averaged states are more suitable in practice~\cite{Jarzyna2012}.

When the state of one mode is the coherent state ($|\alpha\rangle$), Caves~\cite{Caves1981}
found that by injecting the squeezed vacuum state, $|\xi\rangle=e^{\frac{1}{2}(\xi^{*}a^2
-\xi a^{\dagger 2})}|0\rangle$ (here $|0\rangle$ refers to the vacuum state and
$\xi$ the squeezing parameter), in the other mode, the standard quantum limit
can be surpassed. Lang and Caves~\cite{Lang2013} further investigated
this problem and showed that the squeezed vacuum state is actually optimal. It
is shown by expressing the QFI in terms of the variance of the momentum as
$F_{\phi}=2|\alpha|^2\langle\Delta^2 \hat{p}\rangle+N_b$, where
$\langle\Delta^2 \hat{p}\rangle:=\langle\hat{p}^2\rangle-\langle\hat{p}\rangle^2$
with $\hat{p}=\frac{1}{\sqrt{2}i}(b-b^{\dagger})$\,\footnote{The Planck units
($\hbar=1$) is applied here. The other mode is for the coherent state $|\alpha\rangle$.}
and $\langle\cdot\rangle:=\mathrm{Tr}(\cdot\rho)$ denotes the expected value, $N_b$
is the photon number in the second mode. It can be seen that the maximal $F_{\phi}$
is obtained when the variance of the momentum in the second mode is maximal. By
utilizing the inequalities
\begin{equation}
\langle \Delta^2\hat{p}\rangle+\langle\Delta^2\hat{x}\rangle\leq
\langle\hat{p}\rangle^2+\langle\hat{x}\rangle^2=2N_b+1,
\end{equation}
\begin{equation}
\left(\langle\Delta^2\hat{p}\rangle-\langle\Delta^2
\hat{x}\rangle \right)^2\leq 4N_b(N_b+1),
\end{equation}
where $\hat{x}=\frac{1}{\sqrt{2}}(b+b^{\dagger})$ is the position operator of the second mode,
one can obtain
\begin{equation}
2\langle\Delta^2\hat{p}\rangle\leq 2\sqrt{N_b(N_b+1)}+2N_b+1,
\end{equation}
which leads to the maximum QFI
\begin{equation}
F_{\phi, \max}=2N_aN_b+N_a+N_b+2N_a\sqrt{N_b(N_b+1)}.
\end{equation}
This maximal QFI can be achieved with the squeezed vacuum state.
Hence, when one import is fixed to be the coherent state, the squeezed vacuum state is
optimal for the other port. However, the optimal state may not be unique. The existence and
the properties of other optimal states need further investigation.

Lang and Caves~\cite{Lang2014} also studied the optimal state when the input state
of the other import is fixed to be the squeezed vacuum state. Interestingly, the optimal
state in this case is also a squeezed vacuum state. These two squeezed states need to have
opposite values of $\xi$ to satisfy the phase-matching condition.

In the case of the Bayesian estimation where the priori information of the phase is a
flat distribution, it is also possible to perform the state optimization analytically. For
instance, Berry and Wiseman~\cite{Berry2000,Berry2001} studied the state optimization in
the MZI~\cite{Sanders1995}. The target function they used is the Holevo phase
variance~\cite{Holevo1984} $S^{-2}-1$ with
\begin{equation}
S=\left|\int^{\pi}_{-\pi}p(\phi)e^{i\phi}\mathrm{d}\phi\right|
\label{eq:sharpness}
\end{equation}
representing the sharpness of the phase distribution.
For a fixed photon number, the optimal input
state, which maximizes $S$, is given by~\cite{Berry2000,Berry2001}
\begin{equation}
\sqrt{\frac{2}{N+2}}\sum^{N/2}_{m=-N/2}\sin\left(\frac{(2m+N+2)\pi}{2(N+2)}\right)|m\rangle,
\label{eq:adapt_optstate}
\end{equation}
where $|m\rangle$ is the eigenstate of $J_y$ for a fixed $N$ with the eigenvalue $m$.

Apart from the direct optimization, the convex optimization has also been widely used due
to the convexity of the QFI as~\cite{Fujiwara2001,Toth2014}
\begin{equation}
F_x(a\rho_1+(1-a)\rho_2)\leq a F_x(\rho_1)+(1-a)F_x(\rho_2)
\end{equation}
with $a\in [0, 1]$. The convexity of the QFI indicates that the maximal QFI can always be attained
by a  pure state. Since a mixed state can always be expressed as a convex combination of pure states as
$\rho=\sum_i p_i|\psi_i\rangle\langle\psi_i|$ with $\sum_i p_i=1$, we then have have
\begin{eqnarray}
F_x[\mathcal{E}_x(\rho)] &\leq & \sum_{i}p_iF_x[\mathcal{E}(|\psi_i\rangle\langle\psi_i|)] \nonumber \\
&\leq & \max_{i}F_x[\mathcal{E}_x(|\psi_i\rangle\langle\psi_i|)].
\end{eqnarray}
With this convexity, Takeoka et al.~\cite{Takeoka2017} proved a no-go theorem
for the MZI. This no-go theorem states that with the unknown phase shifts in both arms of the
MZI, which is the case for the gravitational wave detection, the scaling of the precision cannot
go beyond the standard quantum limit if the input of one import is vacuum.

Although the convexity of the QFI implies the optimal state is pure in general, in some specific
scenarios the optimization within the mixed states could still be meaningful. In 2011 Modi et
al.~\cite{Modi2011} studied the parameter estimation in terms of quantum circuits in which the
initial states are prepared via the Hadamard and Control-Not gates from mixed
qubit states. The state optimization in this case is performed directly from the
analytical expression of the QFI. Fiderer, Fra\"{i}sse and Braun~\cite{Fiderer2019} considered
the unitary parameterization where the initial state is prepared via unitary operations on a fixed
mixed state, namely, $\rho_{\mathrm{in}}=U\rho_{\mathrm{fix}}U^{\dagger}$ with $U$ as any unitary
operator and $\rho_{\mathrm{fix}}$ a fixed mixed state. The state then goes through a unitary parametrization
as $\rho=U_x\rho_{\mathrm{in}}U^{\dagger}_x$, where $U_x$ is the unitary operator encodes the parameter.
In this case, the state optimization is restricted to the state space that has the same spectrum as
$\rho_{\mathrm{fix}}$. Denote the spectral decomposition of $\rho_{\mathrm{fix}}$
as $\rho_{\mathrm{fix}}=\sum^{d}_{i=1}\lambda_i|\lambda_i\rangle\langle\lambda_i|$
with $d$ the dimension of the density matrix, and $\lambda_i$ is ordered as
$\lambda_1\geq\dots\geq\lambda_d$. Utilizing the operator $\mathcal{H}$, the
optimal initial state is proved to be in the form $\rho_{\mathrm{in,opt}}=
\sum^{d}_{i=1} \lambda_i|\phi_i\rangle\langle\phi_i|$ with
\begin{equation}
|\phi_i\rangle=\begin{cases}
\frac{1}{\sqrt{2}}\left(|\mathcal{H}_i\rangle+|\mathcal{H}_{d-i+1}\rangle\right),
& \mathrm{if}~2i<d+1,\\
|\mathcal{H}_i\rangle, & \mathrm{if}~2i=d+1, \\
\frac{1}{\sqrt{2}}\left(|\mathcal{H}_i\rangle-|\mathcal{H}_{d-i+1}\rangle\right),
& \mathrm{if}~2i>d+1.
\end{cases}
\end{equation}
Here $|\mathcal{H}_{i}\rangle$ is an eigenstate of $\mathcal{H}$ with $\mathcal{H}$ the generator
Hamiltonian of $U_x$ defined as
\begin{equation}
\mathcal{H}=i\left(\partial_x U^{\dagger}_x\right) U_x
=-i U^{\dagger}_x \left(\partial_x U_x\right).
\label{eq:Hgenerator}
\end{equation}
$\mathcal{H}_i|\mathcal{H}_{i}\rangle=h_i|\mathcal{H}_{i}\rangle$ with the eigenvalue, $h_i$, arranged
in decreasing ordered as $h_1\geq\dots\geq h_d$, then the maximal QFI is
\begin{equation}
F_{x,\mathrm{opt}}=\frac{1}{2}\sum^d_{i=1}c_{i,d-i+1}\left(h_i-h_{d-i+1}\right)^2,
\end{equation}
where $c_{i,j}=0$ if $\lambda_i=\lambda_j=0$ and $(\lambda_i-\lambda_j)^2/
(\lambda_i+\lambda_j)$ for others.

In the case that the generator Hamiltonian of $U_x$ is time-dependent and controllable,
i.e., $H=H_x(t)+H_{\mathrm{c}}(t)$, where $H_x(t)$ depends on the unknown parameter
$x$ and $H_{\mathrm{c}}(t)$ is the control Hamiltonian.
They further found an upper bound of
the QFI (denoted by $F_{\mathrm{up}}$),
\begin{equation}
F_{\mathrm{up}}=\frac{1}{2}\sum^d_{i=1}c_{i,d-i+1}
\left[\int^t_0 \left(\mu_i-\mu_{d-i+1}\right)\mathrm{d}\tau\right]^2,
\end{equation}
where $\mu_i=\mu_i(t)$ is an eigenvalue of $\partial_x H_x(t)$ with corresponding
eigenstate $|\mu_i(t)\rangle$ with $\mu_i$ arranged in decreasing ordered, $\mu_1\geq\dots\geq\mu_d$.
The optimal initial state that attains this upper bound is $\rho_{\mathrm{in,opt}}
=\sum^d_{i=1}\lambda_i|\phi^{\prime}_i\rangle\langle\phi^{\prime}_i|$, where
\begin{equation}
|\phi^{\prime}_i\rangle=\begin{cases}
\frac{1}{\sqrt{2}}\left(|\mu_i(0)\rangle+|\mu_{d-i+1}(0)\rangle\right),
& \!\!\mathrm{if}~2i<d+1, \\
|\mu_i(0)\rangle, & \!\!\mathrm{if}~2i=d+1, \\
\frac{1}{\sqrt{2}}\left(|\mu_i(0)\rangle-|\mu_{d-i+1}(0)\rangle\right),
& \!\!\mathrm{if}~2i>d+1.
\end{cases}
\end{equation}

Correa et al.~\cite{Correa2015} studied the optimal probe state for fully thermalized
thermometers, which are in equilibrium state with the reservoir. Assuming the probe is a
$N$-dimensional system, they found the optimal thermalized state, $\rho = e^{-\beta H}/Z$,
requires the Hamiltonian to be an effective two-level system with a highly degenerate
excited state and a specific temperature-dependent gap, here $\beta=k_{\mathrm{B}}T$
with $k_{\mathrm{B}}$ as the Boltzmann constant and $T$ as the temperature. Denote
the Hamiltonian as $H=\sum^{N-1}_{i=0}E_i|E_i\rangle\langle E_i|$, then the optimal
energy structure needs to satisfy
\begin{equation}
(E_i-E_j)\left[E_i+E_j-2(\langle H\rangle+T)\right]=0,
\end{equation}
for any $i$ and $j$ with $\langle H\rangle=\frac{1}{Z}\sum^{N-1}_{i=0}E_i e^{-\beta E_i}$,
which is equivalent to $E_i=E_j$ or $E_i+E_j=2(\langle H\rangle+T)$. This condition
implies that the optimal structure is a degenerate two-level system with the energy
gap $2(\langle H\rangle+T)$. The temperature dependence of the optimal energy gap
means a tunable degenerate two-level system is required in this scheme and the
measurement has to be performed adaptively.

The analytical form may also be identified when the state is restricted to be Gaussian.
In 2006, Monras~\cite{Monras2006} studied the case of a single mode Guassian
state under the unitary channel, $\exp(-ixa^{\dagger}a)$. They found that
the squeezed vacuum state is optimal in this case. \u{S}afr\'{a}nek and
Fuentes~\cite{Dominik2016} further considered several specific cases including the
estimation of a unitary channel that combines the phase-change, squeezing, as well
as the generalized two-mode squeezing and mode-mixing channels. They provided the
optimal states by the direct calculation of the QFI with analytical optimizations.

Knysh et al.~\cite{Knysh2014} developed a method to identify the optimal probe
states for noisy dynamics in the asymptotic limit, i.e., with an asymptotically
large number of qubits or photons. They mapped the problem of identifying the
optimal states to that of finding the ground state of a quantum-mechanical
particle in a 1D potential, where the form of the potential can be determined
from the type of the noise. With this method, they identified the optimal probe states for
various noisy dynamics, including both individual and collective dephasing,
relaxation and excitation, as well as combinations of the above noises. For typical noisy
dynamics the optimal probes are found to approach a Gaussian profile in the
asymptotic limit~\cite{Knysh2014}.

\subsection{Semi-analytical optimization}

The search of optimal states for noisy dynamics is much more difficult in general.
Most noisy dynamics are described in terms of the differential equations and in many
cases they cannot be solved analytically, which makes the analytical state optimization
impossible. In the case that the analytical solutions of the density matrix, or the QFI,
can be obtained, the feasibility of the analytical optimization is still not promised.
Nevertheless, some analytical expressions are definitely useful for the optimization,
which leads to the semi-analytical optimization.

Due to the convexity of the QFI, many methods in the convex optimization have been
successfully applied in the state optimization. It has been applied to the lossy MZI
illustrated in Fig.~\ref{fig:MZI}(b), where the photon loss in the optical MZI is
modeled by the fictitious beam splitter. Dorner, Demkowicz-Dobrzanski et
al.~\cite{Dorner2009,Rafal2009} studied the state optimization in a lossy MZI where
one or two arms suffer the photon losses, as illustrated in Fig.~\ref{fig:MZI}(b).
The input state, $|\psi_{\mathrm{in}}\rangle$, is taken as a two-mode pure state
with fixed total photon number ($N$),
\begin{equation}
|\psi_{\mathrm{in}}\rangle=\sum^{N}_{k=0}c_k|k,N-k\rangle, \label{eq:input_MZI}
\end{equation}
where $|k,N-k\rangle$ is the two-mode Fock state and $c_k$ is a complex coefficient.
In the case that only one arm suffers the photon loss, the QFI can be obtained
directly, which is shown to be a concave function of $\{|c_k|^2\}$. This means
that the local maximum of the QFI is also the global maximum. The optimization
can be written as
\begin{eqnarray}
& \min & -F_x(\{|c_k|^2\}) \nonumber \\
& \mathrm{s.t.} & |c_k|^2\geq 0.
\end{eqnarray}
The interior-point method with the logarithmic barrier
function is then used to numerically locate the optimal state~\cite{Forsgren2002}.
The results show that in the case of $N=10$, the optimal state for a small loss
rate is a N00N-type state $c_0|0N\rangle+c_N|N0\rangle$, which, when the
loss rate is 0, reduces to the N00N state, $\frac{1}{\sqrt{2}}(|0N\rangle+|N0\rangle)$.
It is also found that the performance of the state $c_k|k,N-k\rangle+c_N|N0\rangle$ with
the optimal coefficient $c_k$ and optimal $k$ is very close to the optimal state in a
wide regime of the loss rate.

When equal photon losses exist in both arms, the QFI is not easy to obtain.
Instead, an upper bound of the QFI, $F_x(\sum_i p_i|\psi_i\rangle
\langle\psi_i|)\leq\sum_i p_iF_x(|\psi_i\rangle\langle\psi_i|)$, has been used as the target function
for the state optimization. Again by utilizing the interior-point methods, it is
found that the N00N-type state is optimal for $N<10$. In the case that $N=10$, the
standard N00N state is optimal for small losses and with the increase of the loss
rate the performance of the state~\cite{Berry2000,Berry2001,Maccone2009}
\begin{equation}
\sqrt{\frac{2}{N+2}}\sum^N_{k=0}\sin\!\left(\frac{(k+1)\pi}{N+2}\right)\!|k,N-k\rangle
\end{equation}
is very close to the optimal one. Knysh et al.~\cite{Knysh2011} also studied the state
optimization in the lossy MZI with the input state described in Eq.~(\ref{eq:input_MZI}).
They considered the case where only one arm has photon loss and derived an upper bound on the QFI
as $4(1-R)N/R$, where $R$ is the reflectivity of the fictitious beam splitter.
The optimal state is then obtained by minimizing the difference between the upper
bound and the QFI.

\subsection{Numerical optimization}

\begin{figure*}[tp]
\centering\includegraphics[width=18cm]{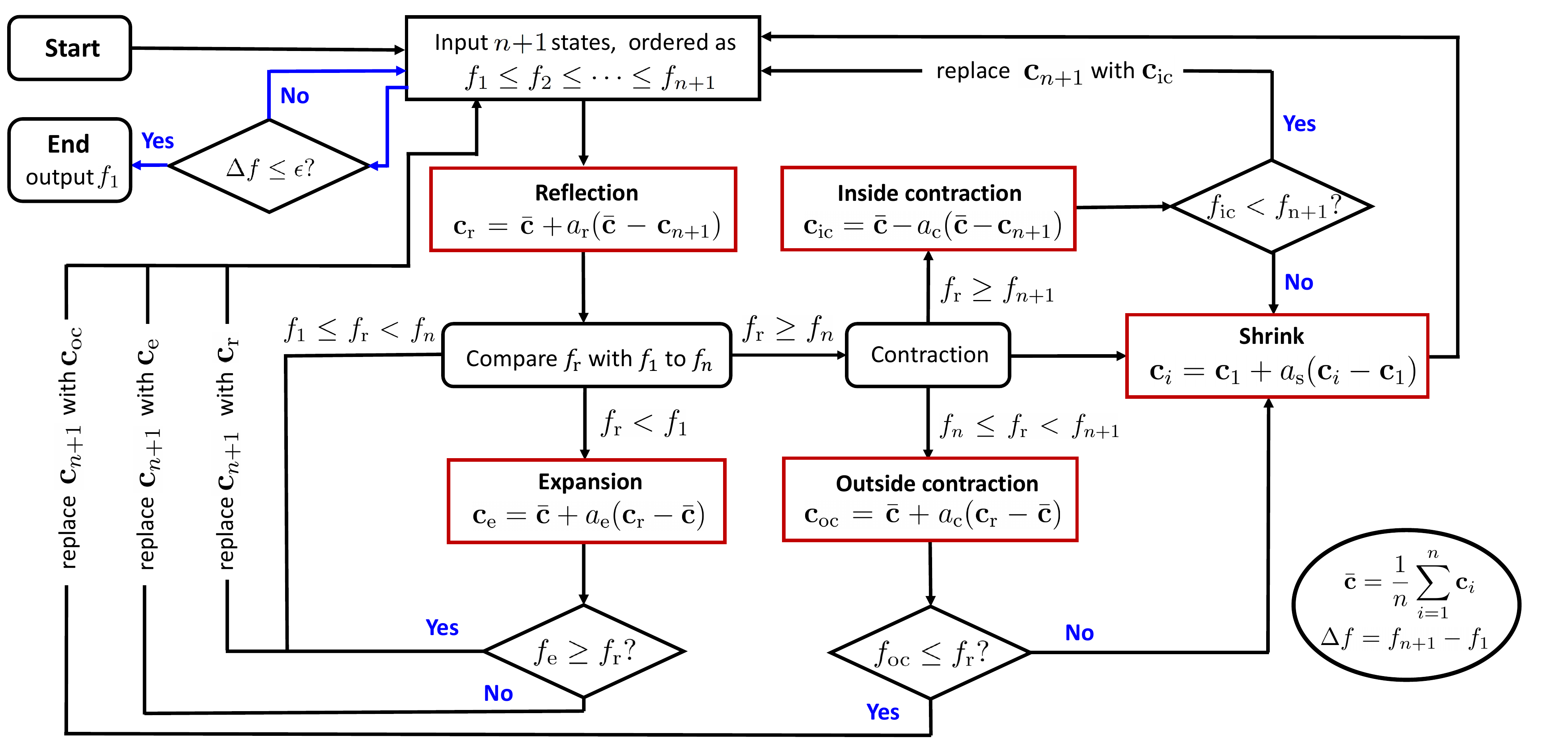}
\caption{Flow chart for the Nelder-Mead algorithm in the state
optimization.}
\label{fig:NM_flowchart}
\end{figure*}

Compared to the analytical optimization, the numerical approaches can be applied to
more general scenarios. Although most searching algorithms can be used in the state optimization,
the algorithms typically can only be applied to systems with limited sizes due to the curse of
the dimensionality.One way to solve this problem is to manually reduce the search space. For
example, Fr\"{o}wis et al.~\cite{Frowis2014} has chosen a subspace to perform the optimization.
They considered the frequency estimation of a collective spin system suffering the
dephasing noise. The system Hamiltonian is $H=S_z:=\frac{1}{2}\sum^N_{i=1}\sigma^{(i)}_z$,
where $\sigma^{(i)}_z$ is the Pauli Z matrix for the $i$th spin and $N$ is the number
of spins. Both the local and collective dephasing have been studied. To reduce the
search space, the states are restricted to the form
\begin{equation}
|\psi\rangle=\sum^{J}_{m=-J}c_m|J,m\rangle,
\end{equation}
where $|J, m\rangle$ is the Dicke state with $J=N/2$. Such state can be characterized by
$\bold{c}=(c_{-J},\cdots,c_J)$. With this ansatz, the search space
reduces to the state space with the maximum angular momentum $J$. It can be further
reduced by assuming $c_m$ is real and positive, and $c_m=c_{-m}$ as
the dynamics is unchanged under the collective spin flipping, $\sigma_x^{\otimes N}$.

Within this subspace, Fr\"{o}wis et al. applied the Nelder-Mead algorithm~\cite{Nelder1965}
to perform the state optimization that maximizes $F_x/t$. The Nelder-Mead algorithm
is a gradient-free search method. The flow chart of the Nelder-Mead algorithm to locate
the minimum target function in the case of the state optimization
is shown in Fig.~\ref{fig:NM_flowchart}, which includes the operations of the reflection,
expansion, contraction and shrink. The first step of this algorithm is to take $n+1$
states $\{\bold{c}_1,\cdots, \bold{c}_{n+1}\}$ and calculate the objective function
\begin{equation}
f_i=-\frac{1}{T}F_x(\bold{c}_i,T)
\end{equation}
at a given time $T$, order them as $f_1\leq f_2\cdots\leq f_{n+1}$.
Next, calculate the reflection state $\bold{c}_{\mathrm{r}}=\bar{\bold{c}}
+a_{\mathrm{r}}(\bar{\bold{c}}-\bold{c}_{n+1})$, here $a_{\mathrm{r}}>0$ is the
reflection coefficient and $\bar{\bold{c}}=\frac{1}{n}\sum^{n}_{i=1}\bold{c}_i$, which leads
to an updated objective function $f_{\mathrm{r}}$. The states are then updated
based on the relation between $f_{\mathrm{r}}$ and the values of the existing
objective functions as following:
\begin{itemize}
\item if $f_1\leq f_{\mathrm{r}}<f_{n}$, replace $\bold{c}_{n+1}$ with
$\bold{c}_{\mathrm{r}}$ and start over;
\item if $f_{\mathrm{r}}<f_1$, let
$\bold{c}_{\mathrm{e}}=\bar{\bold{c}}+a_{\mathrm{e}}(\bold{c}_{\mathrm{r}}-\bar{\bold{c}})$,
and replace $\bold{c}_{n+1}$ with
$\bold{c}_{\mathrm{r}}$ ($\bold{c}_{\mathrm{e}}$) if $f_{\mathrm{e}}\geq f_{\mathrm{r}}$
($f_{\mathrm{e}}<f_{\mathrm{r}}$). This is called the expansion step.
\item
if $f_n\leq f_{\mathrm{r}}<f_{n+1}$, $\bold{c}_{\mathrm{oc}}=\bar{\bold{c}}
+a_{\mathrm{c}}(\bold{c}_{\mathrm{r}}-\bar{\bold{c}})$, and if $f_{\mathrm{r}}\geq f_{n+1}$
let $\bold{c}_{\mathrm{ic}}=\bar{\bold{c}}-a_{\mathrm{c}}(\bar{\bold{c}}-\bold{c}_{n+1})$.
Replace $\bold{c}_{n+1}$ with $\bold{c}_{\mathrm{ic}}$ (or $\bold{c}_{\mathrm{oc}}$)
if $f_{\mathrm{ic}}<f_{n+1}$ ($f_{\mathrm{oc}}\leq f_{\mathrm{r}}$). This is called the
contraction step. If $f_{\mathrm{ic}}$ or $f_{\mathrm{oc}}$ fails to satisfy the
conditions, then replace all the states as $\bold{c}_i=\bold{c}_1+a_{\mathrm{s}}
(\bold{c}_i-\bold{c}_1)$ and start over.
\end{itemize}
The algorithm stops when $\Delta f=f_{n+1}-f_1<\epsilon$ where $\epsilon$ is a given
precision. In this algorithm, the coefficients $a_{\mathrm{e}}>\max\{1, a_{\mathrm{r}}\}$
and $0<a_{\mathrm{c}}, a_{\mathrm{s}}<1$. A usual set of coefficients is
$a_{\mathrm{r}}=1, a_{\mathrm{e}}=2, a_{\mathrm{c}}=a_{\mathrm{s}}=1/2$.

Utilizing the Nelder-Mead algorithm, Fr\"{o}wis et al.~\cite{Frowis2014} performed
the numerical optimization for the estimation of the frequency up to 70 qubits,
where the dynamics of the system is described by the master equation with the dephasing
\begin{equation}
\partial_t \rho=-i[\omega S_z,\rho]+\frac{\gamma}{2}\sum_{i}(\sigma^{(i)}_z\rho\sigma^{(i)}_z-\rho)
\end{equation}
with $\omega$ the frequency to be estimated.
The numerical result suggests that although
the optimal one-axis twisted spin-squeezed state shows a very good performance,
it is still not exactly optimal up to $N\approx 70$. It indicates that the spin
squeezed state is not necessary optimal for the frequency estimation when the
number of qubits is finite, although asymptotically certain type of the spin
squeezed state is optimal~\cite{Knysh2014,Zhou2020}. At the presence of the collective dephasing
\begin{equation}
\partial_t \rho=-i[\omega S_z,\rho]-\frac{\gamma}{1-e^{-\gamma t}}
\left[S_z,\left[S_z,\rho\right]\right]
\end{equation}
with $\omega$ the frequency to be estimated. The numerical result suggests that the Greenberger-Horne-Zeilinger state
$\frac{1}{\sqrt{2}}(|\frac{N}{2},-\frac{N}{2}\rangle+|\frac{N}{2},\frac{N}{2}\rangle)$
is the optimal state.

In 2016 Knott~\cite{Knott2016} provided a search algorithm  for the state optimization
in quantum optics, which is inspired by the evolutionary algorithms. This algorithm
takes the QFI as the target function. It considers a practical state preparation
scenario with two optical modes. The input states in these modes are first chosen
from several types of well-studied states in quantum optics, including the coherent
state, squeezed vacuum state, and the Fock state, which then goes through a series
of operations and then a heralding measurement is performed in one mode at the end.
The operations includes the beam splitter operator, the displacement operator, a
phase shift in one mode, the identity matrix and a non-unitary operator that measures
the state in one mode and input a new state. The heralding measurement is the
photon-number resolving detection, including $1$ to $4$ photons. The algorithm
first randomly picks the input states, the operations and the heralding measurement
from the toolbox and calculates the corresponding QFI. If it is large,
it is taken as the parent and then create an offspring by making a random change
in the process. If the offspring still provides a large QFI, then it is used to
continue to create the next-generation offspring. Hence, this algorithm is similar
to a random search algorithm. Applied to the case of a MZI with phase shifts on
both arms, this algorithm finds states that outperform the squeezed vacuum state
by a constant factor. However, the scaling is still at the standard quantum limit
due to the no-go theorem given in Ref.~\cite{Takeoka2017}.

Apart from the aforementioned algorithms, many other search and optimization
algorithms can be applied to the state optimization, such as the particle swarm
optimization~\cite{Kennedy1995}, the learning algorithms including the actor-critic
algorithm~\cite{Sutton1999} and deep deterministic policy gradient algorithm~\cite{Lillicrap2015}.
These methods are currently being merged in the QuanEstimation
package\,\footnote{https://github.com/LiuJPhys/QuanEstimation.} which will be
thoroughly discussed in a forthcoming paper.

\section{Optimization of the parameterization processes}
\label{sec:parameterization_opt}

\begin{algorithm}[tp]
\SetArgSty{<texttt>}
\caption{GRAPE algorithm~\cite{Liu2017a}}\label{algorithm:grape}
Initialize the control amplitude $V_k(t)$ for all $t$ and $k$; \\
\For {episode=1, $M$}{
Receive initial state $\rho_1$ (i.e. $\rho_{\mathrm{in}}$); \\
\For {$t=1, T$}{
Evolve with the control $\rho_{t+1}=e^{\Delta t\mathcal{L}_t} \rho_{t}$; \\
Calculate the propagators $\mathcal{D}_{t+1}^{t}=\openone,
\mathcal{D}_{t}^{t}=e^{\Delta t \mathcal{L}_t},
\mathcal{D}_{0}^{t}=\mathcal{D}_{t}^{t}\mathcal{D}_{0}^{t-1}$;\\
\For {$i=1, t$}{
Calculate the propagators
$\mathcal{D}_{t-i}^{t}= \mathcal{D}_{t-i+1}^{t} \mathcal{D}_{t-i}^{t-i}$;}}
{Save all $\rho_t$ and $\mathcal{D}$}; \\
Calculate the SLD $L_x(T)$ and QFI $F(T)$; \\
{\For {$t=1, T$}{
\For {$k=1, P$}{
Calculate the gradient $\frac{\delta F(T)}{\delta V_k(t)}$; \\
Update control $V_k(t)\!\leftarrow\! V_k(t)\!+\!\epsilon\frac{\delta F(T)}
{\delta V_k(t)}$;
}}}}
\end{algorithm}

\begin{algorithm*}[tp]
\SetArgSty{<texttt>}
\caption{A3C algorithm~\cite{Mnih2016}} \label{algorithm:A3C}
Randomly initialize global network parameters $\theta$ and $\omega$; \\
\For {episode=1, $M$}{
Reset gradient: $\mathrm{d}\theta=0$, $\mathrm{d}\omega=0$; \\
Synchronize thread-specific parameters $\theta' \leftarrow \theta$, $\omega'\leftarrow\omega$; \\
Receive initial state $\rho_1$ (i.e. $\rho_{\mathrm{in}}$); \\
\For {$t=1, T$}{
Pick the action $V_t$ from $\pi(\rho_{t};\theta')$; \\
Evolve with the control and receive a reward $r_t$ and state $\rho_{t+1}$; \\
Assign the value function $V(\rho_t;\omega')$ to discount reward $R_t$
($R_t=0$ for the last time step); \\}
Save all $r_t$, $\rho_t$ and $R_t$; \\
{\For {$i=T, 1$}{
Update the discount reward $R_i \leftarrow r_i + \gamma R_i$;\\
Update accumulate gradients $\mathrm{d}\theta \leftarrow \mathrm{d}\theta+
\nabla_{\theta'}\mathrm{log}(\pi(\rho_{i}; \theta'))[R_i-V(\rho_i; \omega')]$,
$\mathrm{d}\omega\leftarrow\mathrm{d}\omega+\nabla_{\omega'}[R_i-V(\rho_i;\omega')]^2$;\\
}}
Update the network parameters $\theta\leftarrow\theta+\mathrm{d}\theta$,
$\omega\leftarrow\omega+\mathrm{d}\omega$.}
\end{algorithm*}

The optimization of parameterization process is a major component for the enhancement
of the precision limit. In this section we will introduce the existing optimization
methods for the parameterization processes.

\subsection{Quantum control}

During the evolution of the parametrization process, quantum controls can be employed to alter the dynamics in a desired
way which can improve the precision limit. The controls during the parametrization can either be discrete pulses or
continuous wave forms.

When the dynamics is unitary with a time independent Hamiltonian, $H(x)$,
one optimal control that gives the maximal QFI is to reverse the free evolution,
i.e., by adding a control Hamiltonian $H_c=-H(x)$~\cite{Yuan2015,Yuan2016}. Since
the value of the parameter is a priori unknown, in practice the controls need to be
designed adaptively as $H_c=H(\hat{x})$ with $\hat{x}$ as the updated estimation from the accumulated
data. Such optimally controlled schemes have been experimentally implemented for the estimation of the
parameters in SU(2) operators~\cite{Hou2019,Hou2021b}. Pang and Jordan~\cite{Pang2017} considered the
unitary dynamics with time-dependent Hamiltonian, $H(x, t)$. It is shown that the QFI for a general
time-dependent unitary parametetization process is upper bounded as
\begin{equation}
\sqrt{F_x(t)}\leq\int^t_0 h_{\max}(\tau)-h_{\min}(\tau)\mathrm{d\tau},
\end{equation}
where $h_{\max(\min)}$ is the maximum (minimum) eigenvalue of $\partial_xH(x,\tau)$.
This upper bound can be attained by preparing the initial state as $\frac{1}{\sqrt{2}}(|h_{\max}(0)\rangle+|h_{\min}(0)\rangle)$
and using the control to keep the state as
$\frac{1}{\sqrt{2}}(|h_{\max}(\tau)\rangle+|h_{\min}(\tau)\rangle)$ during the entire evolution, here $|h_{\max(\min)}(\tau)\rangle$
is the eigenstate of $\partial_x H(x, \tau)$ with respect to the eigenvalue $h_{\max(\min)}(\tau)$.

\begin{figure*}[tp]
\centering\includegraphics[width=18cm]{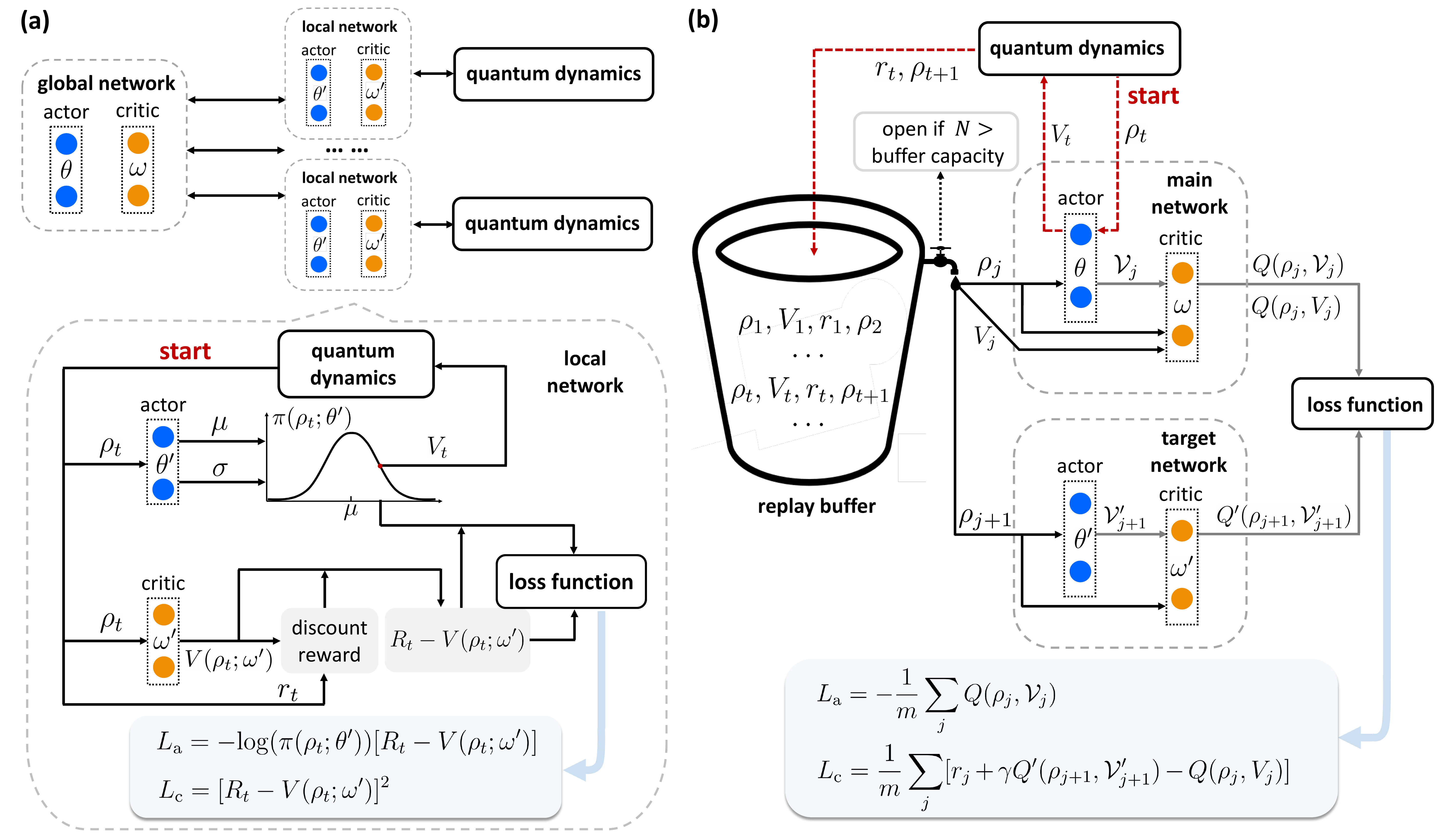}
\caption{(a) Flow chart for the asynchronous advantage actor-critic
algorithm (A3C) in each epoch for the generation of optimal control in quantum
parameter estimation. (b) Flow chart for the deep deterministic policy gradient
algorithm (DDPG) in each epoch for the generation of optimal control in quantum
parameter estimation. (b) is a reprinted figure with permission from~\cite{Tan2021}.
Copyright (2021) by the American Physical Society.}
\label{fig:A3C_DDPG}
\end{figure*}

When the dynamics is noisy, the optimization
in general can only be performed numerically. A widely used optimal
quantum control method is the gradient ascent pulse engineering (GRAPE)
developed by Khaneja et al.~\cite{Khaneja2004}. GRAPE has been employed in quantum metrology under the Markovian noisy dynamics~\cite{Liu2017a,Liu2017b} where the dynamics is described by the master equation
\begin{equation}
\partial_t \rho=-i[H,\rho]+\sum_{k}\left(\!\Gamma_k\rho\Gamma_k^\dagger
-\frac{1}{2}\{\Gamma_k^\dagger\Gamma_k,\rho\}\!\right)\!,
\end{equation}
here the Hamiltonian takes the form
\begin{equation}
H=H_{0}+\sum_k V_k(t)H_k,
\end{equation}
with $H_0$ as the free-running Hamiltonian and $H_k$ as the $k$th control
Hamiltonian with the time-dependent amplitude $V_k(t)$.

For the single-parameter estimation, the steps of the algorithm are presented in
Algorithm~\ref{algorithm:grape}, in which $\Delta t$ is a small time interval,
$\rho_t$ ($\mathcal{L}_t$) is the density matrix  (superoperator) at time $t$.
Before calculating the gradient, a full running of the evolution is required to
obtain the set of propagating superoperators $\mathcal{D}$, which will be used in
the calculation of the gradient. The gradient in this algorithm can be obtained analytically.
The corresponding specific expressions of the gradients and propagators $\mathcal{D}$
can be found in Refs.~\cite{Liu2020,Liu2017a,Liu2017b}, and the codes have been
integrated into the package QuanEstimation. Besides GRAPE, Krotov's method has also
been employed in the design of optimal control in quantum metrology~\cite{Daniel2020}.

When the size of the system increases, the simulation of the dynamics can be difficult.
A hybrid quantum-classical has been proposed to deal with this issue, where the dynamics
is simulated experimentally instead of numerically. Such hybrid approach has been
experimentally demonstrated with the nuclear magnetic resonance~\cite{LiJun2017,Yang2020,Yang2021}.
Other hybrid variants of GRAPE have also been proposed~\cite{Wu2018,Wu2019,Ding2019,Ge2020},
which can be employed in quantum metrology.

Apart from the gradient-based algorithms, learning algorithms have also been employed
for the generation of optimal control in quantum metrology. For example,
Xu et al.~\cite{Xu2019} used the asynchronous advantage actor-critic algorithm (A3C),
a reinforcement learning algorithm, to generate optimal controls in qubit systems.
The pseudocode of A3C algorithm for quantum estimation is given in
Algorithm~\ref{algorithm:A3C} and the corresponding flow chart is given in Fig.~\ref{fig:A3C_DDPG}(a).
This algorithm contains a global actor network with a distribution $\pi(\rho_t; \theta)$
as the output and a global critic network with a value $V(\rho_t; \omega)$ as the
output. Here $\theta$ and $\omega$ are vectors of the network parameters and $\rho_t$
is the density matrix at $t$th time step as well as the input of both networks.
Before training, it assigns the task to multiple threads to enable the parallelism
with local parameters $\theta'$ and $\omega'$ copied from the global networks.
A local actor network picks the action from $\pi(\rho_t; \theta)$ and receives a
reward $r_t$ related to the QFI($F_t$), which can be the QFI itself or
$(F_t-\eta F_{\mathrm{no},t})/F_{\mathrm{no},t}$~\cite{Xu2019} with $\eta\in(0,1]$
and $F_{\mathrm{no},t}$ as the QFI without the control. A local critic
network generates the value function $V(\rho_t; \omega)$, the parameters in the
global network are then updated sequentially after the final ($T$th) time step
according to the accumulate gradients in all local networks.

\begin{algorithm*}[tp]
\SetArgSty{<texttt>}
\caption{DDPG algorithm~\cite{Lillicrap2015}}
Randomly initialize main network parameters $\theta$ and $\omega$; \\
Synchronize target network parameters $\omega'\leftarrow\omega$,
$\theta'\leftarrow\theta$;\\
Initialize replay buffer $\mathcal{R}$;\\
\For {episode=1, $M$}{
Initialize a random distribution $\mathcal N$ for action exploration;\\
Receive initial state $\rho_1$ (i.e. $\rho_{\mathrm{in}}$); \\
\For {$t=1, T$}{
Take the action $V_t=\mu(\rho_{t}; \theta)+\mathcal{N}_t$;\\
Evolve with the control and receive a reward $r_t$ and state $\rho_{t+1}$; \\
Store transition $(\rho_t, V_t, r_t, \rho_{t+1})$ in $\mathcal{R}$;\\
\If {$N>$ buffer capacity}{
Sample a random minibatch of $m$ transitions $\{(\rho_j, V_j, r_j, \rho_{j+1})\}$
from $\mathcal{R}$; \\
Calculate all $y_j=r_j+\gamma Q'(\rho_{j+1}, \mathcal{V}'_{j+1};\omega')$
in the minibatch; \\
Calculte the gradient $\nabla_{\theta}L_{\mathrm{a}}$ with
$L_{\mathrm{a}}=-\frac{1}{m}\sum_jQ(\rho_j, \mathcal{V}_j; \omega)$; \\
Calculate the gradient $\nabla_{\omega}L_{\mathrm{c}}$ with
$L_{\mathrm{c}}=\frac{1}{m}\sum_j[y_j-Q(\rho_j, V_j; \omega)]^2$; \\
Update main network parameters $\theta\leftarrow\theta+\nabla_{\theta}L_{\mathrm{a}}$,
$\omega\leftarrow\omega+\nabla_\omega L_{\mathrm{c}}$;\\
Update target network parameters
$\theta'\leftarrow\tau\theta +(1-\tau)\theta'$,
$\omega'\leftarrow\tau\omega +(1-\tau)\omega'$ with $\tau$ a small weight.
}}}
\label{algorithm:DDPG}
\end{algorithm*}

The performance of A3C algorithm works well in qubit systems, especially combined with
the proximal policy optimization algorithm~\cite{Xu2019}. However, it becomes
hard to converge when the dimension of the system increases. In these cases, the
deep deterministic policy gradient (DDPG) algorithm might be a better choice, which
has been demonstrated to be feasible in the generation of optimal control for certain
estimations involves multiple parameters~\cite{Xu2021}. It has also been used to
enhance the spin squeezing~\cite{Tan2021}. The pseudocode of DDPG algorithm for quantum
estimation is given in Algorithm~\ref{algorithm:DDPG} and the corresponding flow chart
is given in Fig.~\ref{fig:A3C_DDPG}(b). This algorithm contains two actor-critic
networks which are usually referred to as the main networks and target networks,
respectively. The main actor network outputs a value $\mu(\rho_t; \theta)$, which,
by adding a random distribution $\mathcal{N}_t$, gives the action $V_t$ as
$V_t=\mu(\rho_{t}; \theta)+\mathcal{N}_t$. The main critic network outputs a
value $Q(\rho_t, V_t; \omega)$, which is treated as a loss function for the update of
$\theta$. For the sake of minimizing the correlations between samples, $N$ sets
of data $(\rho_t, V_t, r_t, \rho_{t+1})$ are stored in a replay buffer $\mathcal{R}$,
where $r_t$ is the reward related to the QFI. When the storage is finished, a
random minibatch of $m$ transitions in the buffer are put into the main and
target networks during each time step, which is used to obtain the gradients with
respect to $\theta$ and $\omega$. The main networks are then updated accordingly. The
update of the target networks are typically much slower than the main networks to avoid
sharp waving of the total rewards. Feedback controls can also be used to enhance
the QFI~\cite{Zheng2015,LiuLQ2020}.

For non-Markovian dynamics, recently studies have also obtained the optimal precision which
can be computed with the semi-definite programming~\cite{Yang2019,Anian2021}.

\subsection{Quantum error correction}

Quantum error correction is an important tool to battle the noises. Since the pioneer work by
Peter Shor~\cite{Shor1995}, general theory of quantum error correction has been
developed~\cite{Nielsen2002,Steane1996,Knill1997,Gottesman2010,Cory1998,Chiaverini2004}
and applied in quantum computation~\cite{Knill1996,Preskill1997,Aharonov1999,
Chamberland2016} and quantum communication~\cite{Holevo1998,Schumacher1997,Wilde2013}.
Recently, the error correction has also been applied in quantum metrology for the
enhancement of the precision limit~\cite{Zhou2020,Kessler2014,Arrad2014,Dur2014,Lu2015,
Matsuzaki2017,Sekatski2017,Demkowicz2017,Layden2019,Chen2020,Gorecki2020,Zhou2018}.

Given a general noisy quantum channel
\begin{equation}
\mathcal{E}(\rho_{\mathrm{in}})=\sum^{m}_{j=1}K_j\rho_{\mathrm{in}}K^{\dagger}_j,
\end{equation}
the task of quantum error correction is to find a subspace (or error-correcting code)
$\mathcal{C}$ of the Hilbert space such that for any quantum state $\rho_{\mathrm{c}}$
in the subspace, there always exists a recovery channel $\mathcal{R}$ which can
eliminate the effect of the noise, i.e.,
\begin{equation}
\mathcal{R}\left(\mathcal{E}(\rho_{\mathrm{c}})\right)\propto\rho_{\mathrm{c}},
\end{equation}
where $\rho_{\mathrm{c}}$ is a state in the subspace $\mathcal{C}$. This is possible
if each of the error (Kraus) operators $K_i$ maps the code $\mathcal{C}$ to undeformed
and respectively orthogonal subspace, which is detectable and correctable. This forms
the condition of quantum error correction~\cite{Nielsen2002,Knill1997,Gottesman2010},
specifically, $\mathcal{R}$ exists if and only if all the Kraus operators satisfy
\begin{equation}
\Pi_{\mathrm{c}} K_j^\dagger K_l\Pi_{\mathrm{c}}=c_{jl}\Pi_{\mathrm{c}},
\label{eq:QECcondition}
\end{equation}
where $\Pi_{\mathrm{c}}$ is the projective operator to the code space spanned by
the code $\mathcal{C}$ and $c_{jl}$ is a complex constant satisfying $c_{jl}=c_{lj}^*$.

In quantum metrology, besides correcting the noise, a useful quantum error correction code
must also protect the information of the parameter at the same time. Without loss of generality~\cite{Demkowicz2017,Zhou2018},
assume the Hamiltonian of the dynamics takes the form $H=xG$ with $x$ as the unknown parameter and $G$ as the generator,
and for a small time interval $\mathrm{d}t$ the dynamics is described by $\mathcal{E}(U_{\mathrm{d}t}(\rho_{\mathrm{c}}))$
with $U_{\mathrm{d}t}(\rho_{\mathrm{c}})=e^{-iH\mathrm{d}t}\rho_{\mathrm{c}} e^{iH\mathrm{d}t}$ and
$\mathcal{E}(\rho)=\sum_j K_j\rho K_{j}^{\dagger}$. The goal of error correction is then to design a
recovery operation $\mathcal{R}$ such that $\mathcal{R}(\mathcal{E}(U_{\mathrm{d}t}(\rho_{\mathrm{c}})))$
is effectively unitary in the code subspace $\mathcal{C}$ with nontrival parametrization. Such
recovery operation $\mathcal{R}$ exists if and only if~\cite{Kessler2014}
\begin{eqnarray}
(1) & & \Pi_{\mathrm{c}}K_j^\dagger K_l\Pi_{\mathrm{c}}
=c_{jl}\Pi_{\mathrm{c}},~\forall j,l; \\
(2) & & \max_{|\Psi\rangle\in\mathcal{C}}\langle\Delta G_{\mathrm{eff}}^2\rangle>0, \nonumber
\end{eqnarray}
where $G_{\mathrm{eff}}= \Pi_{\mathrm{c}} G \Pi_{\mathrm{c}}$ is the effective generator in the code space,
$\langle\Delta G_{\mathrm{eff}}^2\rangle=\langle\Psi|G_{\mathrm{eff}}^2|\Psi\rangle-\langle\Psi|G_{\mathrm{eff}}|\Psi\rangle^2$.
Condition (1) straightforwardly comes from Eq.~(\ref{eq:QECcondition}), which ensures the existence of a recovery
operation $\mathcal{R}$. Condition (2) requires the error-corrected dynamics depends nontrivially on the parameter
$x$ and the maximal QFI of $x$ is non-zero.

When the noise is Markovian and the dynamics is described by the master equation as
\begin{equation}
\partial_t \rho=-i[H,\rho]+\sum_{k}\gamma_k\left(\!\Gamma_k\rho\Gamma_k^\dagger
-\frac{1}{2}\{\Gamma_k^\dagger\Gamma_k,\rho\}\!\right)\!,
\end{equation}
The error-correction condition can be equivalently formulated as~\cite{Demkowicz2017,Zhou2018}
\begin{eqnarray}\label{eq:QECformetrology}
(1) & & \Pi_{\mathrm{c}}\Gamma_j\Pi_{\mathrm{c}}=\lambda_j\Pi_{\mathrm{c}},~\forall j; \nonumber \\
(2) & & \Pi_{\mathrm{c}}\Gamma_j^\dagger\Gamma_l\Pi_{\mathrm{c}}=\mu_{jl}\Pi_{\mathrm{c}},~\forall j,l;\\
(3) & & \Pi_{\mathrm{c}} G\Pi_{\mathrm{c}}\neq\kappa\Pi_{\mathrm{c}}. \nonumber
\end{eqnarray}
Here $\lambda_j$, $\mu_{jl}$ are complex numbers and $\kappa$ is a constant.
Under the error-correction, the state in the code space evolves effectively as
$\partial_t\rho=-i[H_{\text{eff}},\rho]$, where $H_{\text{eff}}=xG_{\text{eff}}
=x\Pi_{\mathrm{c}}G\Pi_{\mathrm{c}}$. With this noiseless evolution, the maximal
QFI is achieved by choosing the initial state as $|\psi_{\mathrm{in}}\rangle
=\frac{1}{\sqrt{2}}(|\lambda_{\min}\rangle+|\lambda_{\max}\rangle)$, where
$|\lambda_{\min}\rangle$ and $|\lambda_{\max}\rangle$ are the eigenstates of
$H_{\text{eff}}$ with respect to the minimum and maximum eigenvalues. And the
corresponding QFI is $F_x=t^2(\lambda_{\max}-\lambda_{\min})^2$, indicating the
Heisenberg limit is achieved. Denote the space spanned by the Lindblad operators as
$\mathcal{S}=\mathrm{span}\{\bold{I},\Gamma_j,\Gamma_j^\dagger,\Gamma_j^\dagger\Gamma_l\}$,
it has been shown that the precision can reach the Heisenberg scaling if and only
if $H\notin\mathcal{S}$, i.e., the HNLS (Hamiltonian-not-in-Lindblad-span)
condition~\cite{Demkowicz2017,Zhou2018}. When the condition holds, an explicit
construction of the error-correcting code has also been provided by Zhou et al.~\cite{Zhou2018}:
define an inner product between two Hermitian matrices, $A$ and $B$, as $\mathrm{Tr}(AB)$,
the Hamiltonian can then be uniquely decomposed as $H=H_{\parallel}+H_{\bot}$, where $
H_{\parallel}\in\mathcal{S}$ and $H_{\bot}\bot\mathcal{S}$, if $G\notin\mathcal{S}$,
then $H_{\bot}\bot\mathcal{S}$ is nonzero and can be written as $H_{\bot}=\frac{1}{2}
\|H_{\bot}\|_1(\rho_0-\rho_1)$, where $\rho_0$ and $\rho_1$ are trace-one positive
matrices with orthogonal support. By introducing an ancillary system,
$\mathcal{H}_{\mathrm{A}}$, $\rho_0$ and $\rho_1$ can be purified as $|C_0\rangle$
and $|C_1\rangle$ which have orthogonal support in $\mathcal{H}_{\mathrm{A}}$, the
error correction code can then be chosen as the subspace of $\mathcal{H}_{\mathrm{S}}
\otimes\mathcal{H}_{\mathrm{A}}$ spanned by $|C_0\rangle$ and $|C_1\rangle$.

In the quantum error correction protocol for quantum metrology, the degrees of
freedom available for optimization is the error correction code. In the case of
single-parameter estimation, the code optimization is to optimize
$\mathrm{span}\{|C_0\rangle,|C_1\rangle\}$ to let the Hamiltonian has the
maximum gap between the minimum and maximum eigenvalues. Using $\lambda_{\max}-
\lambda_{\min}=\mathrm{Tr}(H_{\mathrm{eff}}\tilde{C})=\mathrm{Tr}(H_{\bot}\tilde{C})$
with $\tilde{C}=\rho_0-\rho_1$, the optimization can be formulated as~\cite{Zhou2018}
\begin{eqnarray}
& \max &\mathrm{Tr}(\tilde{C}H_{\bot}) \nonumber \\
& \mathrm{s.t.} &\Vert\tilde{C}\Vert_{\mathrm{op}}\le 2,  \nonumber \\
& & \mathrm{Tr}(\tilde{C}S)=0,\forall S\in\mathcal{S}.
\end{eqnarray}
Its Lagrange dual problem can be solved efficiently via semidefinite programming as
\begin{eqnarray}
&\min & s \nonumber \\
&\mathrm{s.t.} &
\left(\begin{array}{cc}
s\bold{I} & H_{1}\\
H_1 & s\bold{I}
\end{array}\right)\geq 0
\end{eqnarray}
for variables $\nu_k\in\mathbb{R}$ and $s\ge 0$, here $H_1=H_{\bot}+\sum_k \nu_kE_k$
and $\{E_k\}$ is any basis of $\mathcal{S}$. In the multiparameter case, the code
optimization is much more complicated as one needs to optimize not only the error
correction code but also the input state and the final measurement. An algorithm
is given by G\'{o}recki et al.~\cite{Gorecki2020}  to tackle these problems. For
the dynamics with specific noise, there also exist ancilla-free protocols to achieve
the Heisenberg limit~\cite{Layden2019}. If the HNLS condition is violated, the precision
can only achieve the standard quantum limit, and with approximate quantum error correction,
it has been shown that the upper bound in Eq.~(\ref{eq:Nupperbound}), where only the first
term survives, can be saturated for asymptotically large $N$ up to an arbitrarily small
error~\cite{Zhou2020}. This determines the ultimate precision limit that can be achieved
when the HNLS condition fails.

\section{Optimization of the measurement}
\label{sec:measurement_opt}

Theoretically for the single-parameter estimation the quantum Cram\'{e}r-Rao bound
can be saturated with the projective measurement on the eigenstates of the SLD operator.
However, when the size of the system increases, the eigenstates of the SLD can be hard
to identify. Such measurement can also be highly nonlocal and practically challenging.
In general these measurements can also depend on the value of the unknown parameter,
thus can only be realized adaptively. This often requires a large amount of data processing
that needs to be optimized. Here we review some optimization techniques that are employed
for the local adaptive measurement, which are practically less demanding.

\begin{figure}[tp]
\centering\includegraphics[width=9cm]{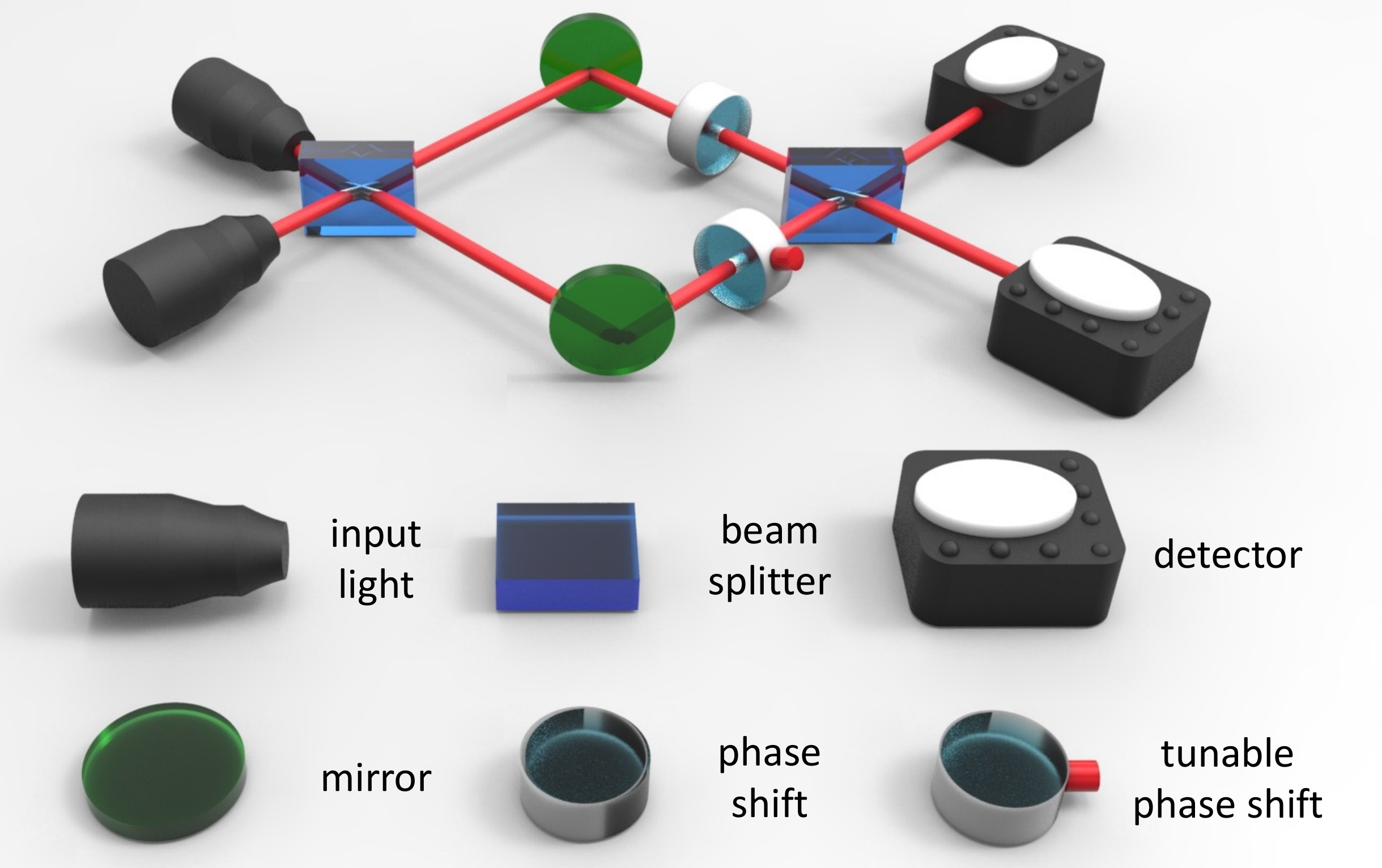}
\caption{Adaptive measurement scheme in MZI. In this scheme,
apart from the unknown phase shift, a known tunable phase
shift exists in other arm.}
\label{fig:adapt_MZI}
\end{figure}

The adaptive measurement was first used in the optical MZI, where apart from the unknown
phase shift, an additional tunable phase is introduced in the other arm in an adaptive way,
as illustrated in Fig.~\ref{fig:adapt_MZI}. It requires many rounds of the measurement
and the tunable phase needs to be adjusted in each round according to the accumulated
measurement data. The major requirements of the adaptive scheme is that the precision
of the known phase has to be much better than that of the phase to be estimated. In this
scheme, how to tune the known phase becomes an important question as it affects both the
precision of the results as well as the efficiency of the scheme.

\begin{figure}[tp]
\centering\includegraphics[width=9cm]{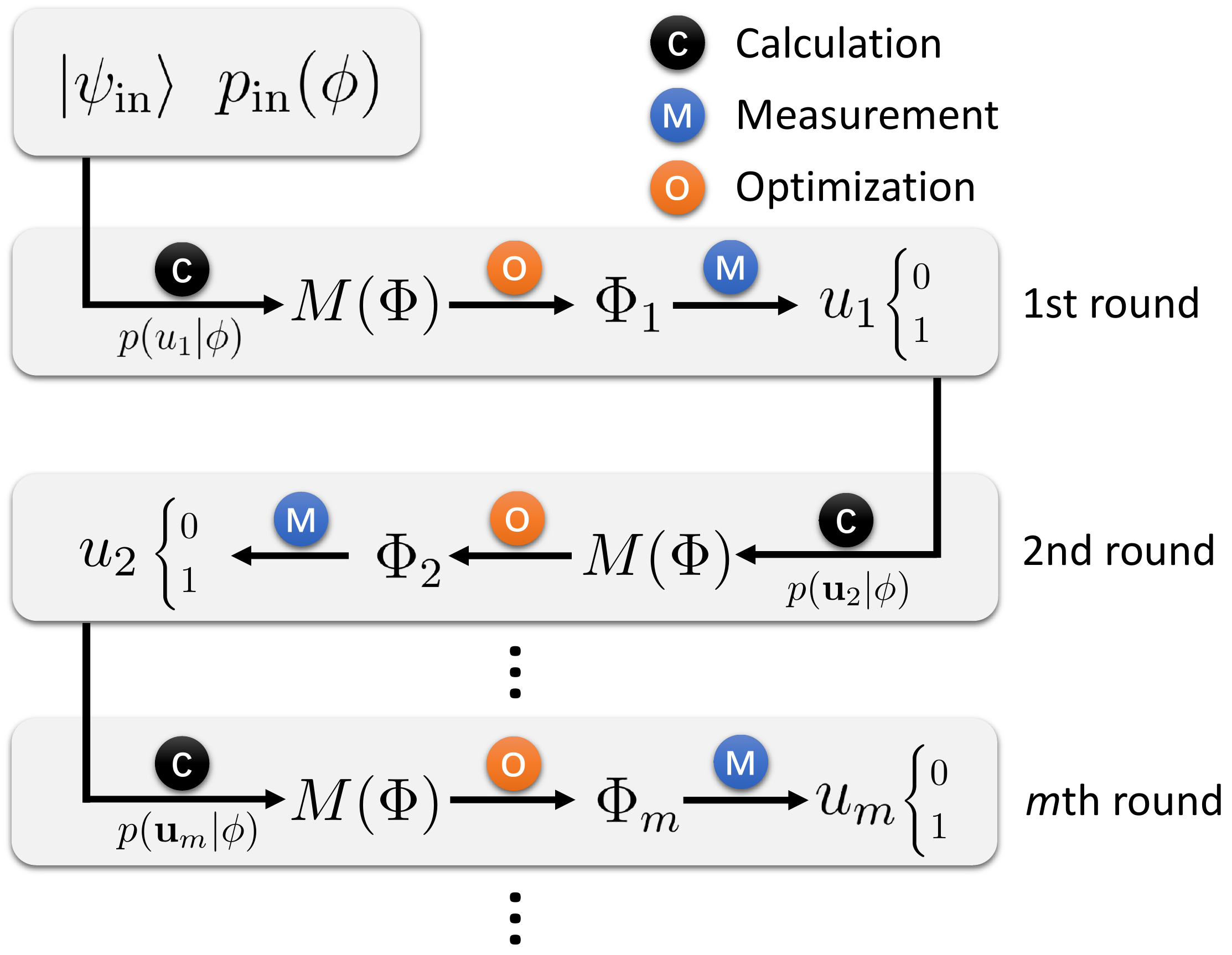}
\caption{Illustration of the process of an online adaptive
measurement. In each round of the measurement, one needs to calculate
the target function $M(\Phi)$ based on $p(\bold{u}_m|\phi)$
and then optimize it to obtain an optimal tunable phase $\Phi_m$,
which is used to adjust the MZI. The measurement is performed with
the adjusted MZI and the result is recorded. For an offline scheme,
no optimization is applied during the experiment and all $\{\Phi_N\}$
is obtained in advance.}
\label{fig:adapt_measurement}
\end{figure}

Both online and offline optimizations have been proposed to optimize the tunabe phase.
In the online approach the tunable phase is adjusted through the real-time optimization
based on the results of the previous rounds, and in the offline approach the adjustment
of the tunable phase is given by a fixed formula which is optimized before the experiment.

Berry et al.~\cite{Higgins2007,Berry2000,Berry2001} proposed an adaptive local measurement
for the phase estimation, where the target function to maximize is taken as the sharpness
function which will be specified below. Different target functions can be employed which
typically converge to the variance in the asymptotic limit. In this scheme, the probe state
$|\psi_{\mathrm{in}}\rangle$ is prepared in the form of Eq.~(\ref{eq:adapt_optstate}) with
the $N$ photons. Then only one photon goes thorough the MZI in each round. The protocol is
illustrated in Fig.~\ref{fig:adapt_measurement}. In the $m$th round, the target function,
which is based on the conditional probability, is calculated which is used to obtain an
optimal tunable phase, $\Phi_m$. We first show the calculation of the conditional probability
given a series of measurement results. The annihilation operator in the output port
of the MZI can be expressed as~\cite{Berry2000,Berry2001}
\begin{eqnarray}
a_u&=&a_{\mathrm{in}}\sin\left(\frac{1}{2}(\phi-\Phi)+\frac{\pi}{2}u\right) \nonumber\\
& &+b_{\mathrm{in}}\cos\left(\frac{1}{2}(\phi-\Phi)+\frac{\pi}{2}u\right),
\label{eq:annihilation_adapt}
\end{eqnarray}
where $a_u$ ($u=0,1$) is the annihilation operator for one output port of the MZI,
$a_{\mathrm{in}}$ and $b_{\mathrm{in}}$ are the annihilation operators for the
input modes, $\phi$ is the unknown phase and $\Phi$ is the tunable phase.
The probability of observing a photon in $a_{0}$
mode, $p(0|\phi)$, in the first round is then
\begin{equation}
p(0|\phi)=\frac{1}{N}\langle\psi_{\mathrm{in}}|a^{\dagger}_{0}a_{0}
|\psi_{\mathrm{in}}\rangle,
\end{equation}
here $1/N$ is the probability of picking one photon from the $N$ photons
and $\langle\psi_{\mathrm{in}}|a^{\dagger}_{0}a_{0}|\psi_{\mathrm{in}}\rangle$
is the probability of detecting this photon in $a_{0}$ mode. Similarly,
$p(1|\phi)=\frac{1}{N}\langle\psi_{\mathrm{in}}|a^{\dagger}_{1}a_{1}|\psi_{\mathrm{in}}\rangle$.
When the measurement result has been recorded in this round, the post-measurement
state becomes
\begin{equation}
|\psi_{u_1}\rangle=a_{u_1}|\psi_{\mathrm{in}}\rangle,
\end{equation}
where $u_1=0,1$ is the result of the first round, i.e., $u_1=0$ ($u_1=1$) means
a photon is detected in $a_0$ ($a_1$) mode. Physically, the post-measurement
state becomes a $(N-1)$-photon state as one photon has been absorbed during the measurement,
which is described by the annihilation operator on $|\psi_{\mathrm{in}}\rangle$. Note
that $|\psi_{u_1}\rangle$ is \emph{unnormalized}, the corresponding normalized state reads
\begin{eqnarray}
|\tilde{\psi}_{u_1}\rangle &=& \frac{1}{\sqrt{\langle\psi_{u_1}|\psi_{u_1}\rangle}}
a_{u_1}|\psi_{\mathrm{in}}\rangle \nonumber \\
&=& \frac{1}{\sqrt{\langle\psi_{\mathrm{in}}|a^{\dagger}_{u_1}a_{u_1}
|\psi_{\mathrm{in}}\rangle}}a_{u_1}|\psi_{\mathrm{in}}\rangle.
\end{eqnarray}
In the next (second) round, the probability of observing a photon in $a_{u_2}$
($u_2=0,1$) is
\begin{eqnarray}
p(u_2|\phi)&=&\frac{1}{N-1}\langle\tilde{\psi}_{u_1}|a^{\dagger}_{u_2}a_{u_2}
|\tilde{\psi}_{u_1}\rangle \nonumber \\
&=&\frac{1}{N-1}\frac{\langle\psi_{\mathrm{in}}|a^{\dagger}_{u_1}a^{\dagger}_{u_2}
a_{u_2}a_{u_1}|\psi_{\mathrm{in}}\rangle}{\langle\psi_{\mathrm{in}}|a^{\dagger}_{u_1}
a_{u_1}|\psi_{\mathrm{in}}\rangle},
\end{eqnarray}
where $1/(N-1)$ is the probability of picking a photon from the left $N-1$
photons. The post-measurement state can be similarly obtained.

Repeating this process, in the $m$th round the conditional probability
$p(u_m|\phi)$ can be expressed as
\begin{eqnarray}
& & p(u_m|\phi) \nonumber \\
&=&\frac{1}{N-(m-1)}\langle\tilde{\psi}_{u_{m-1}}|
a^{\dagger}_{u_m}a_{u_m}|\tilde{\psi}_{u_{m-1}}\rangle \nonumber \\
&=&\frac{1}{N-(m-1)}\frac{\langle\psi_{\mathrm{in}}|a^{\dagger}_{\bold{u}_{m}}
a_{\bold{u}_{m}}|\psi_{\mathrm{in}}\rangle}{\langle\psi_{\mathrm{in}}|
a^{\dagger}_{\bold{u}_{m-1}}a_{\bold{u}_{m-1}}|\psi_{\mathrm{in}}\rangle},
\end{eqnarray}
where $a_{\bold{u}_{m}}:=\prod^{m}_{i=1}a_{u_{i}}$. The conditional probability,
denoted as $p(\bold{u}_m|\phi)$, of observing a series of results,
$\bold{u}_{m}=(u_1,u_2,\dots,u_{m})$, is then
\begin{eqnarray}
p(\bold{u}_m|\phi)&=&\prod^{m}_{i=1}p(u_i|\phi) \nonumber \\
&=& \frac{(N-m)!}{N!}\langle\psi_{\mathrm{in}}|a^{\dagger}_{\bold{u}_{m}}
a_{\bold{u}_{m}}|\psi_{\mathrm{in}}\rangle.
\label{eq:adapt_pm}
\end{eqnarray}

\begin{algorithm*}[tp]
\SetArgSty{<texttt>}
\caption{PSO algorithm~\cite{Hentschel2010,Eberhart1995}}
Initialize $\Phi^{(i)}_1$ and $\Delta\Phi^{(i)}_1$ for each $i\in[1,L]$; \\
Initialize all $\varphi^{(i)}_1=0$; \\
\For {$m=1, M$}{
\For {$i=1, L$}{
Receive the adjustments of the feedback phase $\Delta\Phi_m^{(i)}$; \\
Calculate the objective function $M_{\mathrm{off}}(\{\Delta\Phi_m^{(i)}\})$; \\
Calculate the personal best adjustments of the feedback phase
$\Delta\Phi_{m,\mathrm{pb}}^{(i)}=\arg\left(\underset{k\in[1,m]}{\max}
\,M_{\mathrm{off}}\left(\{\Delta\Phi^{(i)}_k\}\right)\right)$;}
Compare all $M_{\mathrm{off}}(\{\Delta\Phi^{(i)}_{m,\mathrm{pb}}\})$ and
determine the global best adjustments of the feedback phase
$\Delta\Phi_{m,\mathrm{gb}}=\arg\left(\underset{i\in[1,L]}{\max}\,M_{\mathrm{off}}
\left(\{\Delta\Phi_{m,\mathrm{pb}}^{(i)}\}\right)\right)$; \\
\For {$i=1, L$}{
Calcluate $\varphi_m^{(i)}=c_0\varphi_{m-1}^{(i)}
+\mathrm{rand}()\cdot c_1(\Delta\Phi_{m,\mathrm{pb}}^{(i)}-\Delta\Phi_m^{(i)})
+\mathrm{rand}()\cdot c_2(\Delta\Phi_{m,\mathrm{gb}}-\Delta\Phi_m^{(i)})$; \\
Update the adjustments of the feedback phase $\Delta\Phi_{m+1}^{(i)}=
\Delta\Phi_m^{(i)}+ \varphi_m^{(i)}$.
}}
\label{algorithm:pso}
\end{algorithm*}

Based on the conditional probability, the tunable phase, $\Phi_m$, can be adjusted.
Suppose $m-1$ rounds of adaptive measurements have been carried out with the result
$\bold{u}_{m-1}$. In~\cite{Berry2000,Berry2001}, the tunable phase in the $m$th round,
$\Phi_m$, is obtained by maximizing the quantity
\begin{equation}
M_{\mathrm{on}}(\Phi)=\sum_{u_m=0,1}p(u_m)S_{u_m},
\label{eq:M_foropt}
\end{equation}
here $S_{u_m}$ is the sharpness function,
\begin{eqnarray}
S_{u_m}&=&\left|\int^{\pi}_{-\pi}p(\phi|\bold{u}_{m})e^{i\phi}\mathrm{d}\phi\right|
\end{eqnarray}
and $p(u_m)=\int^{\pi}_{-\pi}p(u_m|\phi)
p(\phi)\mathrm{d}\phi$ is the probability of observing $u_m$ in the $m$th
round, here $p(\phi)$ is the probability distribution conditional on the recorded
measurement data, $\bold{u}_{m-1}$. Alternatively we can write $p(u_m)=p(u_m|\bold{u}_{m-1})$,
since it is an online scheme and all the probabilities in this round should be based
on the fact that all the results in previous rounds, i.e., $\bold{u}_{m-1}$, have
already been recorded. Therefore,
\begin{equation}
p(u_m)=\frac{p(\bold{u}_m)}{p(\bold{u}_{m-1})},
\label{eq:adapt_pum}
\end{equation}
with
\begin{equation}
p(\bold{u}_m)=\int^{\pi}_{-\pi}p(\bold{u}_{m}|\phi)p_{\mathrm{in}}(\phi)
\mathrm{d}\phi
\label{eq:adapt_pumdef}
\end{equation}
as the probability of observing the result $\bold{u}_m$. Here $p(\bold{u}_m|\phi)$
can be calculated as in Eq.~(\ref{eq:adapt_pm}) and $p_{\mathrm{in}}(\phi)$ is the
original priori probability before the experiment. If no priori information
exists, it can just be chosen as the uniform distribution, i.e., $p_{\mathrm{in}}(\phi)=1/(2\pi)$.

The conditional probability, $p(\phi|u_m)$, in $S_{u_m}$  can be obtained via the
Bayes' rule,
\begin{equation}
p(\phi|\bold{u}_m)=\frac{p(\bold{u}_m|\phi)p_{\mathrm{in}}(\phi)}{p(\bold{u}_m)}.
\end{equation}
The sharpness function then reads
\begin{eqnarray}
S_{u_m}&=&\left|\int^{\pi}_{-\pi}p(\phi|\bold{u}_{m})e^{i\phi}\mathrm{d}\phi\right|
\nonumber \\
&=&\frac{1}{p(\bold{u}_m)}\left|\int^{\pi}_{-\pi}p(\bold{u}_m|\phi)p_{\mathrm{in}}(\phi)
e^{i\phi}\mathrm{d}\phi\right|.
\end{eqnarray}
Thus
\begin{equation}
M_{\mathrm{on}}(\Phi)=\frac{\sum_{u_m}\left|\int^{\pi}_{-\pi}p(\bold{u}_m|\phi)
p_{\mathrm{in}}(\phi)e^{i\phi}\mathrm{d}\phi\right|}{\int^{\pi}_{-\pi}
p(\bold{u}_{m-1}|\phi)p_{\mathrm{in}}(\phi)\mathrm{d}\phi}.
\label{eq:online_M}
\end{equation}
In the first round, the denominator of the above equation should be set to 1. The
optimization is then performed to obtain the optimal tunable phase, $\Phi_m$, which
is used in the MZI for the $m$-th round. Repeating this approach, a complete online
adaptive policy $\{\Phi_m\}$ can be obtained according to the recorded results in
all rounds.


Different from the online strategy where $\{\Phi_m\}$ are determined through the
real-time optimization, $\{\Phi_m\}$ in the offline strategy are generated by
formulas which are optimized before the experiment. One choice of the target function
in the offline strategy, without any measurement data, is to take the average of
the sharpness function over all possible trajectories of $\bold{u}_m$,
\begin{equation}
M_{\mathrm{off}}(\Phi)=\sum_{\bold{u}_m}\left|\int^{\pi}_{-\pi}p(\bold{u}_m|\phi)
p_{\mathrm{in}}(\phi)e^{i\phi}\mathrm{d}\phi\right|.
\label{eq:adapt_Moff}
\end{equation}
The optimal tunable phase, $\Phi_m$, can then be determined by optimizing
$M_{\mathrm{off}}(\Phi)$, which can be performed offline. Hentschel and Sanders~\cite{Hentschel2010}
provided an offline adaptive measurement scheme based on the particle swarm
optimization (PSO)~\cite{Eberhart1995}. In their scheme, the tunable phase in the
real experiment is updated via the rule
\begin{equation}
\Phi_m=\Phi_{m-1}-(-1)^{u_{m-1}}\Delta\Phi_{m},
\label{eq:offline_phim}
\end{equation}
where $u_{m-1}=0,1$ ($m\geq 2$) is the result obtained in the $m-1$ round
experiment, and $\Phi_1$ is set manually. With
this rule, the goal of the strategy is to provide a set of good
\begin{equation}
\{\Delta\Phi_m\}:=\{\Delta\Phi_1,\Delta\Phi_2,\dots,\Delta\Phi_M\}
\end{equation}
via PSO algorithms, where the flow of the algorithm is given in Algorithm~\ref{algorithm:pso}.
As the number of the trajectories in Eq.~(\ref{eq:adapt_Moff}) grows exponentially $(\sim 2^N)$,
in practise $M_{\mathrm{off}}(\Phi)$ is often approximated by averaging a reasonable
number of sampled trajectories~\cite{Hentschel2011,Palittapongarnpim2017,Palittapongarnpim2019}.
And typically only a polynomial number of trajectories are needed to obtain a good approximation.
Peng and Fan~\cite{Peng2020} further proposed an ansatz that reduces the complexity to $N^4$.
Experimentally, an offline scheme based on the PSO algorithm has been realized by Lumino
et al.~\cite{Lumino2018}.

Other optimization methods, such as the genetic algorithm~\cite{Rambhatla2020} and the
differential evolution (DE) algorithm~\cite{Lovett2013}, have also been used in the offline
adaptive measurement. Compared to the PSO algorithm, the performance of the DE algorithm is
more robust, 
and the DE algorithm also works better with a large photon number~\cite{Palittapongarnpim2019}.

Wiseman~\cite{Wiseman1995} considered the adaptive homodyne measurement scheme for the phase
estimation, which was further discussed with both the semiclassical approach~\cite{Wiseman1997}
and the quantum mechanical approach~\cite{Wiseman1998}. The scheme was also experimentally
demonstrated by Armen et al.~\cite{Armen2002} with coherent states as the input states. In 1996,
D'Ariano et al.~\cite{DAriano1996} discussed the two-quadrature measurement with squeezed states
as the input states. Further in 2009, Olivares et al.~\cite{Olivares2009} pointed out the optimal
squeezing parameter with respect to the true value of the unknown phase (or relative phase in the
adaptive measurement) in the homodyne detection and invoked the Bayesian inference to update the
posterior probability. Bayesian estimation based on continuously monitored environment have also
been studied~\cite{Gammelmark2013,Gammelmark2014,Kiilerich2016,Albarelli2018,Rossi2020,Albarelli2017}.

The Bayesian estimator is asymptotically unbiased and is capable to attain the quantum
Cram\'{e}r-Rao bound in the asymptotic limit. In 2015, Berni et al.~\cite{Berni2015} experimentally
realized the adaptive homodyne measurement with the squeezed vacuum states and Bayesian inference. 
Wheatley et al.~\cite{Wheatley2010} used an adaptive homodyne measurement scheme to estimate a stochastically
varying phase shift on a coherent beam utilizing the quantum smoothing techniques introduced by Tsang~\cite{Tsang2009}.
Wiebe and Granade~\cite{Wiebe2016} proposed the rejection filtering to approximate the Bayesian inference which
can reduce the memory of samplings. Zheng et al.~\cite{Zheng2019} experimentally applied this method to an adaptive
Bayesian measurement for the optical phase estimation. Adaptive Bayesian estimation has also been experimentally
applied to multiparameter estimation by Valeri et al.~\cite{Valeri2020}. Recently, Fiderer et al.~\cite{Fiderer2021}
also used the neural network for the adaptive Bayesian quantum estimation. Joint measurements on conjugate observables
have also been demonstrated~\cite{Steinlechner2013}.

We note that in practice the choices of the measurement can be very limited. For example, in optical systems,
the typical measurements are photon counting, homodyne measurement, displacement measurement where the successful
rate is also limited by the hardware efficiency. In practise, it is important to include the practical constraints
in the optimization.

\section{Summary}
\label{sec:summary}

The optimization of the schemes is crucial to gain quantum
advantages in quantum metrology. Here we reviewed the recent development of the optimizations in the three steps
of the schemes in quantum metrology: preparation (Sec.~\ref{sec:state_opt}), parameterization (Sec.~\ref{sec:parameterization_opt})
and measurement (Sec.~\ref{sec:measurement_opt}). In the process of state preparation, analytical, semi-analytical
and numerical approaches, are reviewed. In the parameterization process, both quantum control and quantum error
correction techniques are presented. In the measurement, the optimizations of the adaptive measurement and other
scenarios are summarized. In practice, typically the optimization of the parameterization process is first performed which identifies the optimal control that leads to the channel with the maximal quantum channel Fisher information, the optimal probe state is then identified according to the obtained channel, finally the optimal measurement is determined based on the output probe state.

A practical challenge for the implementation of the optimal schemes, just as in all other
quantum technologies, is the curse of the dimensionality. Although some techniques presented
can reduce the complexity by restricting to a subspace, this either requires certain symmetries
or loses the global optimality. Practical techniques that can provide optimal schemes for
intermediate number---from dozens to thousands---of particles are highly desired. Another practical
challenge is that the systems in practise may not be well characterized. This can be partially
addressed in terms of the nuisance parameters~\cite{Tsang2020,Suzuki2020} or with a full
multi-parameter quantum estimation~\cite{Liu2020,Albarelli2020,Szczykulska2016}. The tradeoff
between the optimality and the robustness of the schemes, however, requires further studies.
The optimal scheme in the finite regime~\cite{Rubio2019,Tsang2012,Lu2016,Liu2016}, even for the
single-parameter estimation, is still not well-understood and there are plenty of rooms for optimizations.

Finally we note that these techniques are not only useful in quantum metrology, but can
also be adopted in various other applications, such as quantum process tomography, quantum
channel discrimination, variational quantum eigensolver, quantum verification, etc., and could
serve as a bridge among different applications.

\medskip
\textbf{Acknowledgements}  \par
The authors would like to thank Jinfeng Qin, Huai-Ming Yu and Yu-Qian Xu for
helpful discussions. This work was supported by National Natural Science
Foundation of China through Grant No.\,11805073, No.\,12175075 and the Research
Grants Council of Hong Kong through the Grant No.\,14307420.

\medskip
\textbf{Conflict of Interest} \par
The authors declare no conflict of interest.

\medskip
\textbf{Keywords} \par
quantum parameter estimation, quantum Cram\'{e}r-Rao bound, quantum control,
optimization

\medskip

\end{document}